\PassOptionsToPackage{pdfpagelabels=false}{hyperref} 
\documentclass[12pt, doublespace, twocolumn, usenatbib, useAMS]{mnras}

\newcommand\lsim{\mathrel{\rlap{\lower4pt\hbox{\hskip1pt$\sim$}}
        \raise1pt\hbox{$<$}}}
\newcommand\gsim{\mathrel{\rlap{\lower4pt\hbox{\hskip1pt$\sim$}}
        \raise1pt\hbox{$>$}}}

\usepackage{amsmath}
\usepackage{widetext}
\usepackage{times}
\usepackage{graphicx}
\usepackage{verbatim}
\usepackage{tikz}
\usepackage{color}

\usepackage[T1]{fontenc}
\usepackage{aecompl}
\usepackage{calligra}
\DeclareMathAlphabet{\mathcalligra}{T1}{calligra}{m}{n}
\DeclareFontShape{T1}{calligra}{m}{n}{<->s*[2.2]callig15}{}

\def\prd{PRD}
\def\apj{ApJ}                 
\def\apjl{ApJL}               
\def\apjs{ApJS}               
\def\mnras{MNRAS}             
\def\aap{A\&A}                
\def\nat{Nature}              
\def\araa{ARA\&A}             

\def\Cnt{\mathcal{C}}

\def\orb{\rm{orb}}

\def\obs{\rm{obs}}

\def\min{\rm{min}}
\def\max{\rm{max}}

\def\EM{\rm{EM}}
\def\Edd{\rm{Edd}}
\def\gal{\rm{gal}}

\def\Msun{{M_{\odot}}}

\def\Num{\mathcal{N}}
\def\Mag{\mathcal{M}}

\def\Chandra{\textit{Chandra}}
\def\Lynx{\textit{Lynx}}
\def\Athena{\textit{Athena}~}
\def\Axis{\textit{Axis}~}

\begin{document}

\title[Binary Self Lensing]{Detecting Gravitational Self Lensing from Stellar-Mass Binaries Composed of Black Holes or Neutron Stars}
\author[D. J. D'Orazio, R. Di Stefano]{Daniel J. D'Orazio$^1$\thanks{daniel.dorazio@cfa.harvard.edu; rdistefano@cfa.harvard.edu},
 Rosanne Di Stefano$^1$ \\
    $^1$Astronomy Department, Harvard University, 60 Garden Street, Cambridge, MA 02138}

\maketitle
\begin{abstract}

We explore a unique electromagnetic signature of stellar-mass compact-object
binaries long before they are detectable in gravitational waves. We show that
gravitational lensing of light emitting components of a compact-object binary,
by the other binary component, could be detectable in the nearby universe.
This periodic lensing signature could be detected from present and future
X-ray observations, identifying the progenitors of binaries that merge in the
LIGO band, and also unveiling populations that do not merge, thus providing a
tracer of the compact-object binary population in an enigmatic portion of its
life. We argue that periodically repeating lensing flares could be observed
for $\lsim 100$~ks orbital-period  binaries with the future \Lynx~X-ray
mission, possibly concurrent with gravitational wave emission in the LISA
band.  Binaries with longer orbital periods could be more common and be
detectable as single lensing flares, though with reliance on a model for the
flare that can be tested by observations of succeeding flares. Non-detection
of such events, even with existing X-ray observations, will help to constrain
the population of EM bright compact-object binaries.

\end{abstract}


\section{Introduction: Lensing Happens}  

The detection of the merger of eleven compact-object (CO) binaries by the
Laser Interferometer Gravitational-Wave Observatory (LIGO) \citep{GW150914,
GW151226, GW170104, GW170608, GW170814, GW170817, LIGO_O2_COBs:2018} has begun
to shape our understanding of binary black hole (BBH) and binary neutron star
(BNS) demographics \citep[e.g.,][]{LIGO_BBHO1:2016, ProgGW170817,
LIGO_O2_COBs:2018}. While ongoing monitoring of the high-frequency
gravitational-wave (GW) sky with LIGO will continue to enhance our
understanding of these fascinating systems, detection at merger offers only a
snapshot in a binary's lifetime. A snapshot from which we aim to infer the
entire life story of a population of CO-binary systems: the mechanisms that
form the binaries and eventually drive them towards merger. Here we offer a
possible electromagnetic (EM) mechanism with the potential to diversify the
demographic sample of CO binaries to include those at an earlier stage of
evolution: outside of the LIGO and the lower GW-frequency LISA
\citep{LISA:2017} snapshots, prior to strong GW emission.

On the theoretical side, proposed mechanisms for the formation of binaries
that will merge in the LIGO band involve a variety of astrophysical processes
that are still uncertain. The two most highly studied channels for forming
LIGO binaries are isolated evolution of massive binary-star systems in the
`field' via the common-envelope \citep[\textit{e.g.},][]{VossTauris:2003,
Dominik+2012, Dominik+2013, Ivanova+2013, Dominik+2015, Belczynski:2016,
Belczynski+2016Natur} and chemically homogeneous evolution
\citep{deMinkMandel:2016, MandeldeMink:2016, Marchant+2016} scenarios, and
dynamical formation in dense cluster environments
\citep[\textit{e.g.},][]{PZMcM:2000, Banerjee+2010, Tanikawa:2013, Bae+2014,
Rodriguez+2015, Rodriguez+2016, Askar+2017, ParkKim+2017},
or via secular dynamical interactions such as the Kozai-Lidov mechanism
\citep{AntoniniPerets:2012, NaozKL:2016, SilTremKL:2017, AntToonHamers:2017,
RandallXianyu:2018, ZSZ_KL+BS:2019}. With other possible scenarios involving
very massive stars, AGN discs, or primordial black holes \citep[BHs;][]{Fryer+2001, Loeb:2016,
Bird+2016, StoneFAU:2017, Bartos+2017, McKernan+2018, DOrazioLoeb:2018}.

To determine which channels dominate, and to gain insight into the astrophysics
regulating each channel, we must aim to determine in what proportion the
proposed formation and merger mechanisms contribute to the observed CO merger
rate. Achieving this relies on predicting unique imprints of formation
channels on binary properties during an observable portion of their
lifetimes. Such properties include mass and spins at inspiral and merger
\citep{Gerosa+2013, Kesden+2015, Gerosa+2015, RodZevinSpins+2016,
GerosaBerti:2017, Fishbach+2017, LiuLai:2017, RodPN1+2018, RodAntonini:2018},
redshift distributions \citep{RodLoeb:2018, RodPN2+2018}, eccentricities
measured near merger in the LIGO band and also earlier in the binary
evolution, with lower frequency GWs from LISA, as well as comparison of
populations across the LIGO and LISA bands \citep{DOrazioLoeb:2018, SDI:2018,
SDII:2018, DSIII:2018, Gerosa+2019MultiBandGW}. With the exception of recent
work \citep{SamsingTDE+2019, KremerTDE+2019}, the majority of methods for measuring such
binary properties can only be employed near merger, when GW emission is
detectable, or when strong EM transients associated with merger may operate
\citep[\textit{e.g.}, explosions related to NS mergers][]{GW170817}. %

In this work we propose an EM signature of CO binaries that would act as a
tracer of CO binary formation and evolution long before detection is possible in
GWs. We propose an EM tracer caused by binary self-lensing whereby at least
one component of the CO binary is bright and is periodically magnified by its
companion via gravitational lensing. Such an EM signature would be uniquely
periodic while having the benefit of being a magnified source of intrinsic
emission. This unique EM signature would provide another handle on CO-binary
populations in a regime complimentary to GW observations, helping
to vet formation and merger scenarios. It is possible (though less-likely)
that binary self-lensing could also serve as an EM counterpart  to future GW
observations with the space-based LISA.

In Section \ref{S:Scales} we consider the spatial and time scales associated
with lensing, in calculations that apply to any pair of merging COs, whether
they consist of NSs, stellar mass BHs, or supermassive BHs. In this work we
focus on CO binaries that, if they merge, would generate gravitational waves in the
LIGO band. We have considered the case of supermassive BHs in a
previous work \citep{DoDi2018}.

In Section \ref{S:WhichSystems}, we consider the optimal binary parameters for
generating high magnification lensing flares that could be found to repeat
over the course of a single or a few X-ray observations ($\lsim 10^2$~ks). We
further generate mock X-ray light curves for hypothetical nearby sources
accreting at the Eddington limit and observed with \Chandra-like, or
future \Lynx-like X-ray telescopes.

In Section \ref{S:Plausibility} we estimates the number of CO binary systems
that we expect to be EM bright at orbital periods and inclinations that can be
probed by existing and future X-ray observations (as motivated in \S
\ref{S:WhichSystems}). We further consider mechanisms for lighting up the
binary components, namely accretion from supernova fall-back or a 
common-envelope stage, or gas overflow from a tertiary stellar companion. In \S
\ref{S:Discussion} we conclude with a discussion of what we may learn from
lensing observations of CO binaries and in \S \ref{S:Conclusions} we conclude.

\section{Binary Self-Lensing: Spatial and Time Scales}
\label{S:Scales}

We consider a binary consisting of two masses, $M_1$ and $M_2$, on a circular orbit with semimajor axis $a$, mass ratio $q =M_2/M_1\leq1$, and total mass $M =M_1+M_2$.
The Schwarzschild radius associated with the total mass is
\begin{equation}    
R_{\rm S,T} = \frac{2 GM}{c^2}.  
\end{equation}    
A natural unit of time is 
\begin{equation}    
{\cal T}_{\rm T,cross}=\frac{R_{\rm S,T}}{c},    
\end{equation}    
the time required for light to traverse a distance equal to the Schwarzschild
radius associated with $M$.

The semimajor axis may be expressed in units of $R_{\rm S,T}$:
$a=N_a\, R_{\rm S,T}$. The orbital period is
\begin{equation}    
P_{\rm orb} = {\cal T}_{\rm T,cross}\, \Bigg[ 2\, \sqrt{2}\, \pi\, \,    
N^{3/2}_a \Bigg] .
\end{equation}
The time to merger is
\begin{equation}    
\tau_{\rm merge} = {\cal T}_{\rm T,cross}\, \Bigg[ 
\frac{5}{32}\, \frac{(1+q)^2}{q}\, 
\Big(N^4_a-N_{a,\rm merge}^4 \Big)\Bigg]  .
\end{equation}
The ratio between the time to merge and the orbital period is
\begin{equation}   
\frac{\tau_{\rm merge}}{P_{\orb}} = \Big(\frac{5}{64\, \sqrt{2}\, \pi}\Big)\, 
\Big(\frac{(1+q)^2}{q}\Big)\, N^{5/2}_a .
\end{equation}

To incorporate lensing, we consider the Einstein radius, which depends on the
line-of-sight distance, $a_p(t)$, between the lens and source. 
The distance from the observer to the binary may range from
several parsecs to hundreds of Mpc, while the source and lens are in a
compact binary. Hence, because the source-lens distance is much smaller 
than the distance to the binary, we can write the Einstein radius of
each component of the binary, $R_{E,i}$ as follows,   
\begin{equation}   
R_{E,i}= \Bigg(2\, a_p(t)\, R_{S,i}\Bigg)^{1/2} = 
R_{S,T}\, \, \Big(2\, g(t)\, f_i(q)\Big)^{1/2}\, N^{1/2}_a.
\end{equation}
Here, $g(t)$ is needed to include the orbital phase dependence of the line-of-sight separation
between source and lens, and $f_i(q)$ includes
the relationship between $R_{S,i}$ and $R_{S,T}$ \citep[See for example
Eq. (1) of][]{DoDi2018}.

The effects of lensing depend on $u(t)$, the projected distance between the
source of light and the foreground lens mass expressed in units of $R_{\rm
E,i}.$ The magnification of a point source by a point lens is $1.34,$ when
$u=1.$ For much smaller values of $u,$ the magnification is $1/u.$  If the
lensed object is a NS, the light that  is deflected is most likely to come
from near the surface of the neutron star:  $R_{\rm source} = R_{\rm NS}$.  If
the lensed object is a BH, light to be deflected may come from the region near
the innermost stable circular orbit (ISCO) at $R_{\rm source}=3\, R_{\rm
S,j}.$ To avoid finite-source-size effects that lead to lower magnification,
we require $u>(1 + \epsilon) \times  R_{\rm source}/R_{\rm E,i}$,  Thus, the
maximum magnification possible, $\Mag_{\max}$, is
\begin{equation} 
\Mag_{\max}=\frac{R_{\rm E,i}}{R_{\rm source} \, (1 + \epsilon)}=
N^{1/2}_a\, 
\frac{(2\, g(t)\, f(q))^{1/2}\, R_{S,T}}{(1 + \epsilon)\, R_{\rm source}} .
\label{Eq:Amax}
\end{equation} 
This peak magnification occurs when the line-of-sight separation between the
source and lens is at its maximum, and it repeats once per orbital period.

We can estimate the duration of the periodic
lensing events to be $\tau_{\rm ev} = 2R_{E,i}/v_{\rm orb}$,  
\begin{equation} 
\tau_{\rm ev}={\cal T}_{\rm T,cross}\, 4\, 
\Big(g(t)\, f(q)\Big)^{1/2}\, N_a  .
\label{Eq:tev}
\end{equation} 
The duty cycle is the fraction of the orbital period during which an event 
is ongoing,
\begin{equation} 
{\cal D} = 
\Big(\frac{1}{\pi}\Big)\, 
\Big(2N_a\Big)^{-1/2}\, 
\Big(g(t)\, f(q)\Big)^{1/2}.
\end{equation} 
The time during which detectable effects occur may be either longer than or
shorter than $\tau_{\rm ev}$, since depending on the situation, the value of
$u$ needed to produce detectable magnification may be either larger than or
smaller than $u=1.$

We can only detect the effects of lensing if the orbit is well-aligned with
our line of sight. The probability that the alignment is suitable is
approximately equal to
\begin{equation} 
{\cal P} = \frac{R_{E,L}}{a} = N^{-1/2}_a \, 
\Big(2\, g(t)\, f(q)\Big)^{1/2} ,
\label{Eq:Prob1}
\end{equation}
where suitable here means that $u\leq 1$.

The probability that the source is aligned within an inclination
that gives $\Mag_{\max}$ is lower and given, for $R_{\rm source} \ll a$, by
\begin{equation}
{\cal P}_{\Mag_{\max}} \sim \frac{2}{\pi} \frac{R_{\rm source}}{a} = \frac{2}{\pi} \frac{\cal P}{\Mag_{\max}} = \frac{2}{\pi} \frac{(1 + \epsilon)R_{\rm source} }{R_{S, T}} N^{-1}_a \sim N^{-1}_a.
\label{Eq:Probmax}
\end{equation}
For choice of $R_{\rm source}  = 3 R_{S, j}$, which is approximately the NS
radius and the size of the BH ISCO, then ${\cal P}_{A\max} \sim N^{-1}_a$ is
only approximate up to a factor of $(1+q)^{-1}$ for the primary lens and
$(1+1/q)^{-1}$ for the secondary lens.

\begin{figure*}
\begin{center}$
\begin{array}{cc}
\includegraphics[scale=0.4]{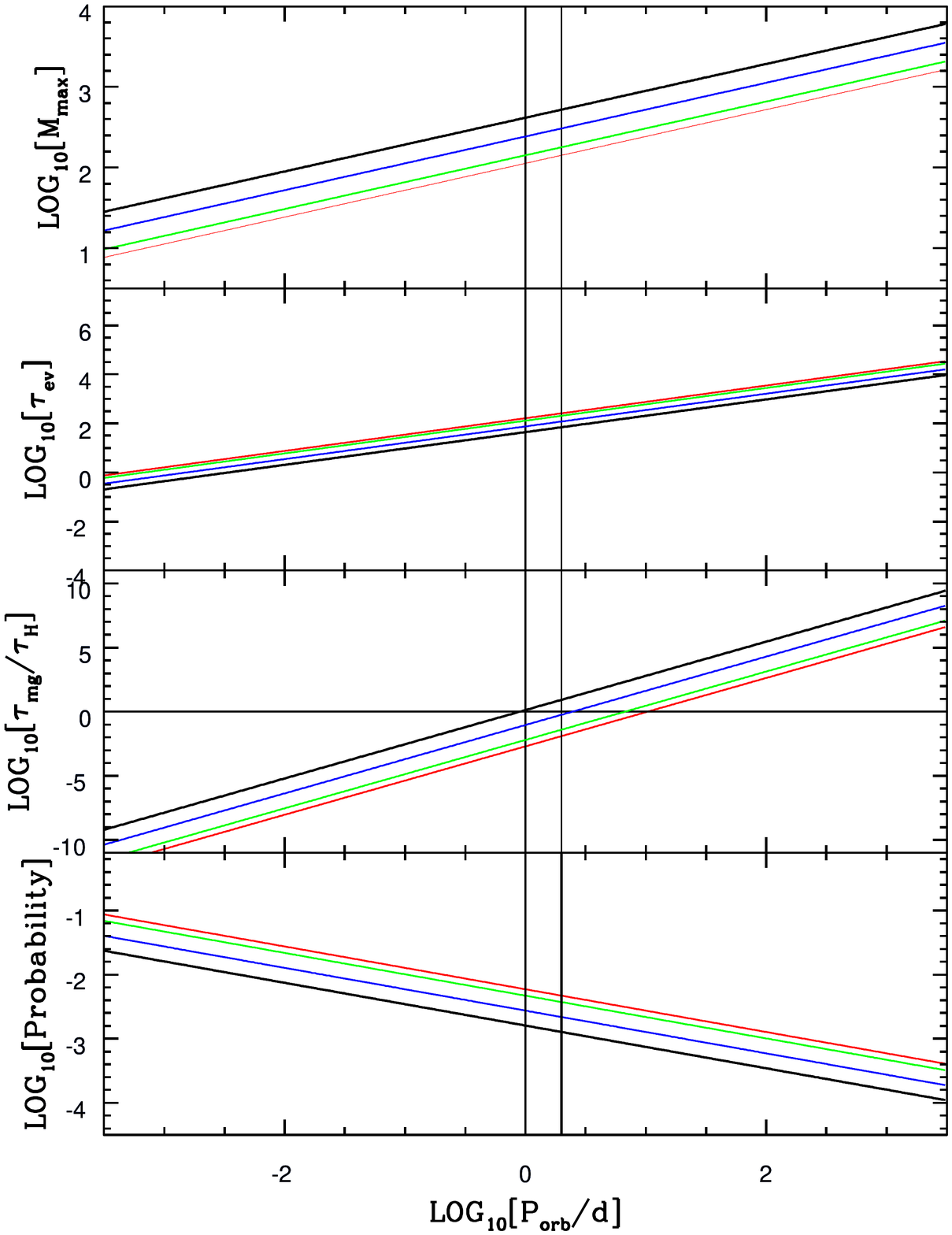} &
\includegraphics[scale=0.4, angle=0]{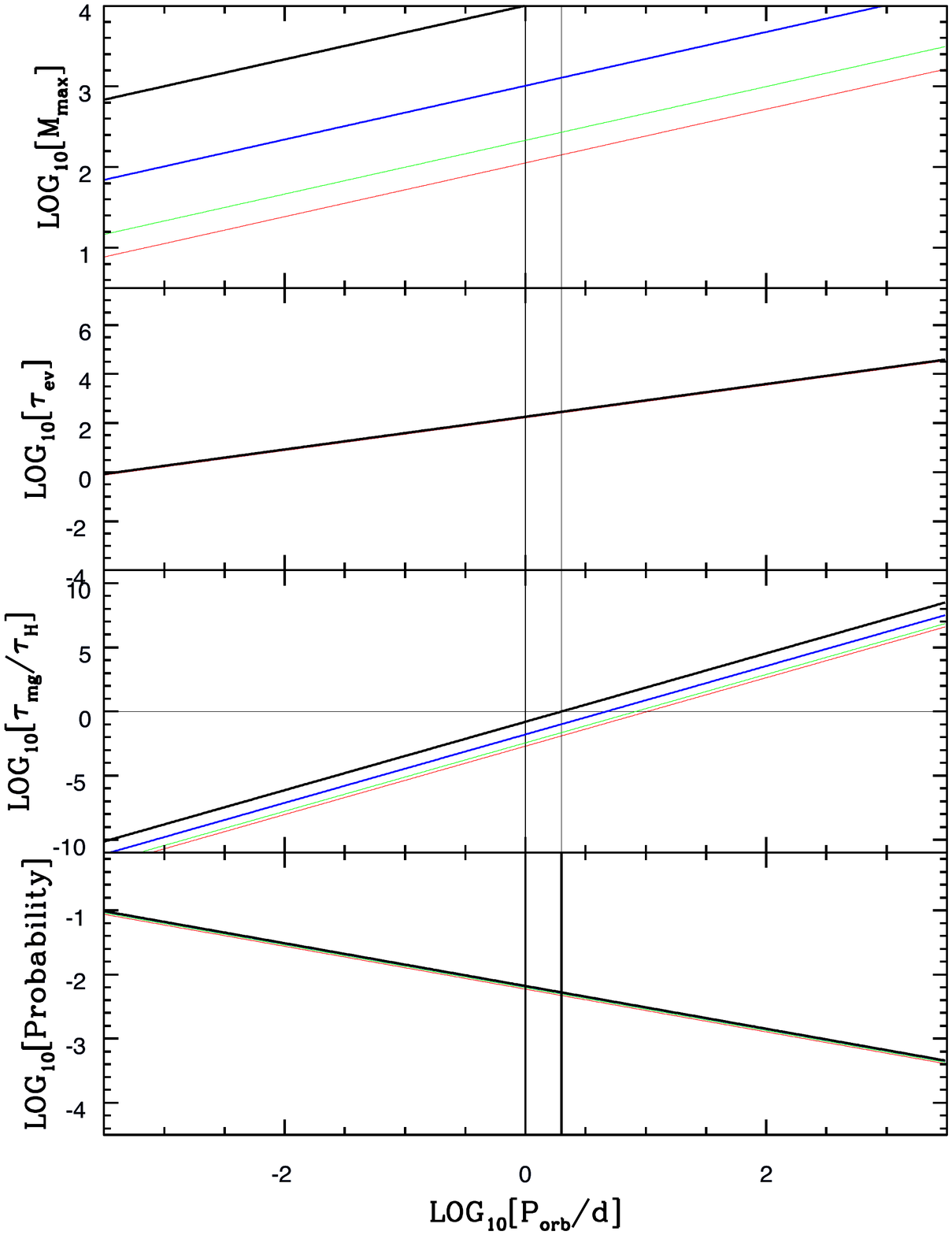}
\end{array}$
\end{center}
\vspace{0pt}
\caption{
Binary self-lensing quantities vs. the binary orbital period in days. From top to
bottom: the maximum possible magnification, the duration of a self-lensing
event in seconds, the time until merger due to gravitational radiation in units of the Hubble time, and the strong-lensing 
probability, Eq (\ref{Eq:Prob1}). In the left panel we
consider four  equal-mass compact-object binaries with total masses $4\Msun$,
$20\Msun$, $100\Msun$ and $200\Msun$ (black, blue, green, and red
respectively, appearing from top to bottom in the top panel). In the right panel we vary the binary mass
ratio for a fixed primary mass of $100 \Msun$ (assumed to be the lens). In
order of appearance from top to bottom in each panel, are the mass ratios,
$q=0.01, 0.02, 0.1, 1.0$ (black, blue, green, and red respectively). Vertical
black lines denote 100~ks and 200~ks orbital periods.
}
\label{Fig:1}
\end{figure*}

In Figure~\ref{Fig:1} we show the relationship between the orbital period and
quantities relevant for detection. For each of four binary masses (left panel)
and four binary mass ratios (right panel), Figure \ref{Fig:1} shows, from top
to bottom: the maximum magnification, the duration of the lensing event,
the time to merger via gravitational wave emission, and the probability for
alignment suitable for strong lensing Eq. (\ref{Eq:Prob1}).

We have selected a region, marked by two parallel vertical lines,
corresponding to orbital periods between  100~ks and 200~ks.  X-ray
observations of external galaxies can cover comparable time intervals, often
in separate exposures, possibly even by different X-ray telescopes.  Some
galaxies within $10-20$~Mpc have been observed for more than 5 times as long.
If the total exposure time is larger than an orbital period there is a good
chance of detecting a lensing spike multiple times. Thus, in the region  
in-between and to the left of the parallel lines, the lensing
hypothesis would be most easily testable through a repeating signature. 

To the right of the parallel lines, a single flare may still be detected and
identified as a lensing candidate, to be followed up with future observations.
These longer orbital-period binaries have longer lasting flares ($\gsim
100$~sec) and higher maximum magnifications ($\gsim 100$), making individual
events more likely to be detected and potentially providing enough counts over
a long enough time interval to allow reliable model fits. These longer orbital
period systems also have a longer GW-driven time to merger (longer than the
Hubble time). Note, however, that if, for example, mass transfer from a star
in a hierarchical orbit is providing the mass that makes one or both compact
objects detectable, mass transfer and mass loss from the system will likely
decrease the time to merger so that some of these systems will merge in a
Hubble time. Finally, the longer orbital period systems have a lower
probability for optimal lensing alignment. In addition, for systems where less
than an entire orbit is observed and the cadence of observation is longer than
the event duration, the complete expression for the probability must include a
factor equal to the observing duty cycle, which is smaller than unity in such
systems. The reason we nevertheless expect that compact-object binaries with
large orbital periods can contribute is that in each galaxy, the numbers of
compact-object binaries with larger orbits and with lower intrinsic
luminosities (that may be detectable with large magnification) is potentially
large (See \S \ref{S:Plausibility}).

Looking at the variation with total binary mass in the left panel of Figure
\ref{Fig:1}, we find that the lower mass systems produce shorter events with
higher maximum magnification. Here the black, blue, green, and red lines
represent equal mass binaries with total masses $4\Msun$, $20\Msun$,
$100\Msun$, $200\Msun$ respectively. The lifetime of such systems under
GW-driven decay is of order a Hubble time for the 100~ks orbital
period, $4\Msun$ binaries. The lifetime drops to $\sim 10^8$ yr for $100\Msun$
binaries with 100~ks orbital periods. Hence, the less massive CO binaries
exhibit shorter lensing durations but longer binary lifetimes and higher peak
magnifications at the orbital periods of interest.

In the right panel of Figure~\ref{Fig:1}, we plot the same quantities, but
each curve corresponds to a different mass ratio. Each binary has a lens
component with $M=100\, M_\odot$, and a second component that is of $1\,
M_\odot, 10\, M_\odot,  50\, M_\odot$, and $100\, M_\odot$ in black, blue,
green, and red, respectively (in the top panel, the uppermost curve is black,
then blue, green, and red, respectively).  We consider the situation in which
the higher mass object lenses the inner disc of the accretion flow onto the
smaller mass. This means that the Einstein radius is relatively large compared
with the X-ray bright region of the disc around the smaller mass. This allows
the maximum magnification to be larger for smaller mass ratio systems. Thus,
the top panel shows the most interesting effect, while the three panels below
illustrate that the rest of the parameters are weakly dependent on the binary
mass ratio. We explore the ideal systems for detection in the next section.

\section{Which Systems can be Detected?}
\label{S:WhichSystems}

\begin{figure*}
\begin{center}
%
\includegraphics[scale=0.5]{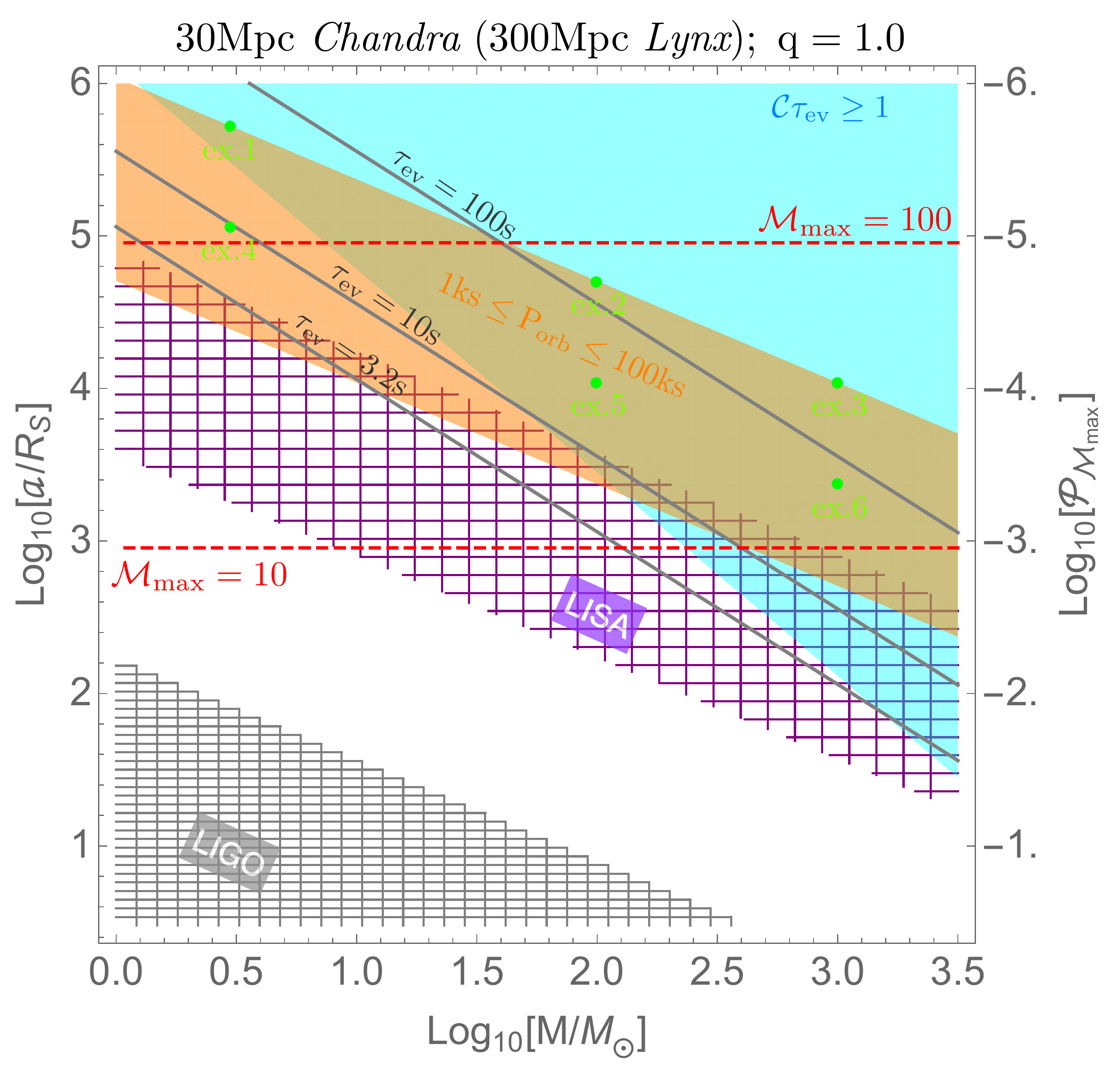}
\end{center}
\vspace{0pt}
\caption{
Regions in binary separation $a$ (in units of Schwarzschild radius of the total binary mass) and total-binary-mass space where lensing flares could be observable with a \Chandra-like (\Lynx-like) X-ray observatory at a distance of 30 Mpc (300 Mpc), assuming sources emit at the Eddington luminosity for their total binary mass. The \textcolor{cyan}{teal region} is where the instrument can detect at least one count per lensing flare duration. The \textcolor{orange}{orange region} delineates binaries with orbital periods between $1-10^2$~ks. The gray lines are contours of constant lensing timescale, the shortest is set to the \Chandra~read-out time. The dashed red lines are lines of constant, maximum lensing magnification (Eq. \ref{Eq:Amax}). The right $y$-axis estimates the probability that a given binary will produce lensing at the maximum magnification, when observed for an entire orbital period. For reference, we draw the approximate regions of sensitivity for GW detection by LISA (GW frequencies of $10^{-3}-10^{-1}$~Hz, hatched purple) and LIGO (GW frequencies $>10$~Hz, hatched gray), assuming a binary on a circular orbit.
}
\label{Fig:2}
\end{figure*}

Figure \ref{Fig:2} demonstrates the range of binary parameters for which
observations are viable assuming an equal mass ratio binary and a source
distance of 30~Mpc for a present-day, \textit{Chandra}-like X-ray telescope
\citep{CSC:2010}, and at 300~Mpc for a future, \textit{Lynx}-like telescope
\citep{Lynx:2018}.

The teal shaded region labeled $\Cnt \tau_{\rm{ev}}\geq1$ estimates
where the magnified number of counts for either telescope would be larger than
one per lensing event duration. This timescale limitation is enforced
to ensure that the lensing flare can be resolved, and observed to repeat.
Longer integration times could of course allow deeper observations, but at the
price of washing out the distinguishing flare. The duration of the lensing event
$\tau_{\rm{ev}}$ is computed from Eq. (\ref{Eq:tev}). The lensed count rate is
given by,
\begin{equation}
\Cnt = \Cnt_0\Mag_{\max},
\label{Eq:CntRate}
\end{equation}
where $\Mag_{\max}$ is given by Eq. (\ref{Eq:Amax}) and the un-lensed count rate
$\Cnt_0$ is found by assuming that the source emits with a bolometric
luminosity at a fraction $f_{\Edd}$ of the Eddington value at a distance $d$,
and that all emission is in the $2-10$ kev range.
Using the NASA HEASEARC WebPIMMS tool for the \Chandra~ACIS-S instrument,
we find,
\begin{equation}
\Cnt_0 \approx  1.3 \ \rm{cnt} \ \rm{s}^{-1} f_{\Edd} \left(\frac{d}{1 \rm{Mpc}}\right)^{-2} \left(\frac{M}{10 \Msun}\right).
\label{Eq:Bkg_cnt}
\end{equation}
where we have adopted a galactic neutral Hydrogen column density of $n_H =
1.22 \times 10^{21}$ cm$^{-2}$ and a photon index of $2.0$
\citep{PhotIndex:Yang+2015}. We then simply scale the count rate with binary
mass and distance.


The orange shaded region in Figure \ref{Fig:2} demonstrates a range of
feasible binary orbital periods for detection of repeating flares
(corresponding to the region to the left of the vertical lines in Figure
\ref{Fig:1}). To maximise the probability of seeing a lensing flare, we would
like the orbital period to be short enough to observe over the course of a few
observations. Hence, we draw a maximum orbital period at $10^2$~ks. For
visualization purposes, we extend the orange region down to a minimum period
of 1~ks, a standard single observation time for X-ray observatories. The lower
limit of the orange shaded region is not a hard cutoff. Rather, there are a
number of factors that affect the minimum observable orbital period for which
observation of a lensing event is possible. We estimate this hard cutoff to be
where the lensing duration is approximately the instrument readout time,
$\approx3.2$ seconds for \Chandra. The gray lines in the left panel of Figure
\ref{Fig:2} are drawn at constant lensing flare duration, including this
instrumental lower limit. The minimum orbital period is also set by the 
count-rate requirement delineated by the teal region, which extends below the 1~ks
period limit for the most massive binaries. Hence, it is a combination of
observed source brightness and instrumental cadence that determine the
shortest orbital period systems for which lensing flares are detectable.

While the shorter orbital-period binaries are likely more rare, provide less
magnification, and exhibit shorter event durations, they do offer a higher
probability of lensing per binary. On the right $y$-axes of the left panel of
Figure \ref{Fig:2} we label the approximate probability of seeing a lensing
event at the maximum magnification (Eq. \ref{Eq:Probmax}). Note that the
probability of seeing a lower magnification flare, given by Eq.
(\ref{Eq:Prob1}), is higher than this and scales with $N^{-1/2}$ rather than
$N^{-1}$. These probabilities of course do not take into account the binary
population as a function of orbital separation and mass. Because closer
separation and higher mass binaries merge more quickly (at least for GW-decay
driven binaries), the probability of seeing a binary at a smaller separation
also goes down, mitigating the higher lensing probabilities. Also, none of the
considerations thus far take into account the time that the binary is EM
bright. We address these issues in \S \ref{S:Plausibility}.

The hatched purple and gray regions in Figure \ref{Fig:2} represent the
approximate location of the LISA (GW frequencies $10^{-3} - 10^{-1}$~Hz) and
LIGO (GW frequencies $>10$~Hz) sensitivity ranges for circular-orbit binaries.
For highly eccentric binaries, such as those that may form in dense stellar
environments, these shaded regions would be shifted upwards because a binary
with high eccentricity, but otherwise the same orbital period, emits the
majority of its GW power at higher harmonics of the binary orbital frequency
than in the circular case. We note that the lensing flare duration will also
change by an amount less than $(1\pm e)$, while the magnification will change
by an amount less than $\sqrt{1\pm e}$ for binary eccentricity $e$, depending
on whether the lensing flare occurs closer to apocentre or pericentre. The
periodic nature and symmetric nature of the lensing flare will not change for
eccentric systems \citep{Spikey:2019}.

\begin{figure*}
\begin{center}$
\begin{array}{ccc}
\includegraphics[scale=0.25]{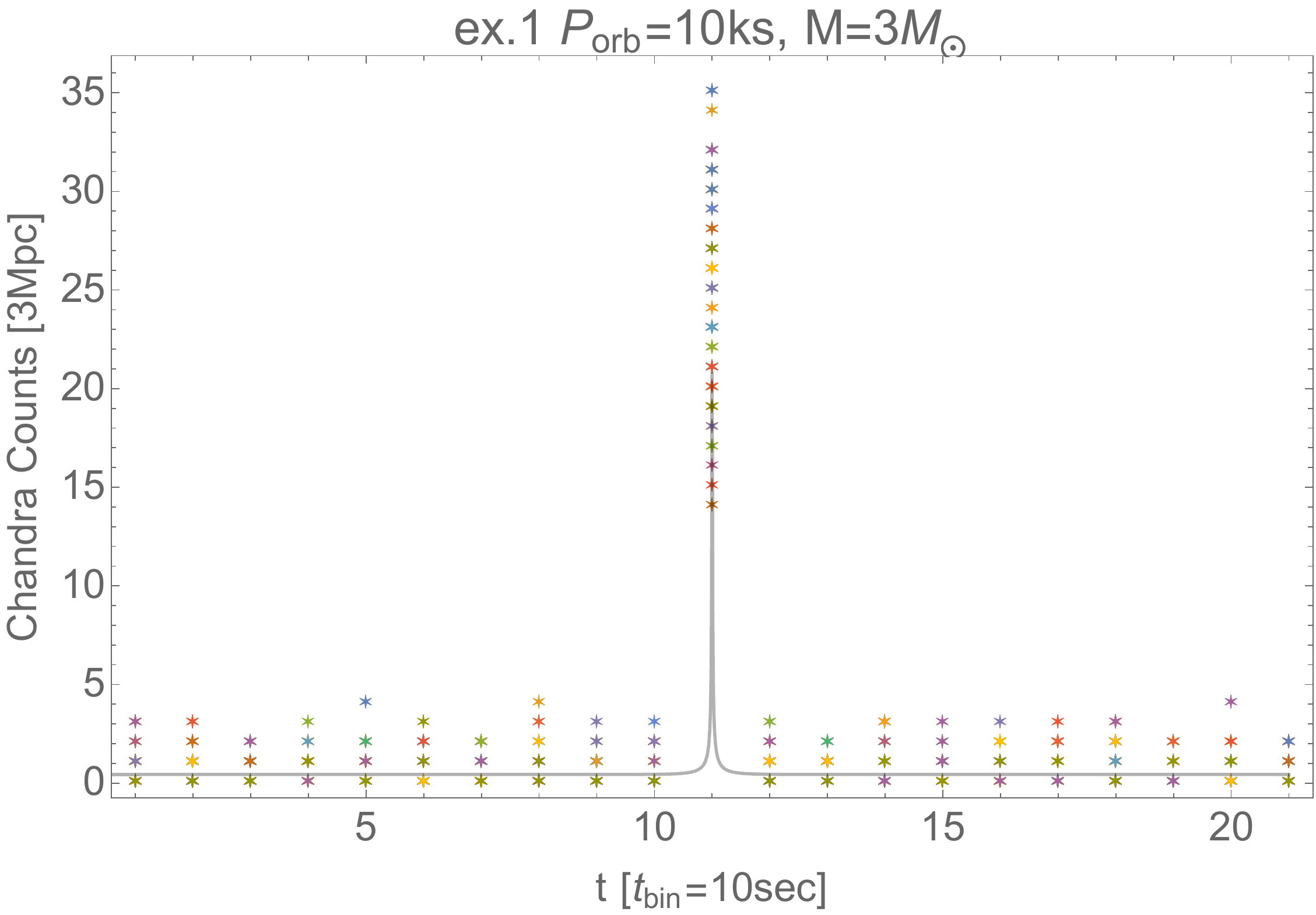} &
\includegraphics[scale=0.25]{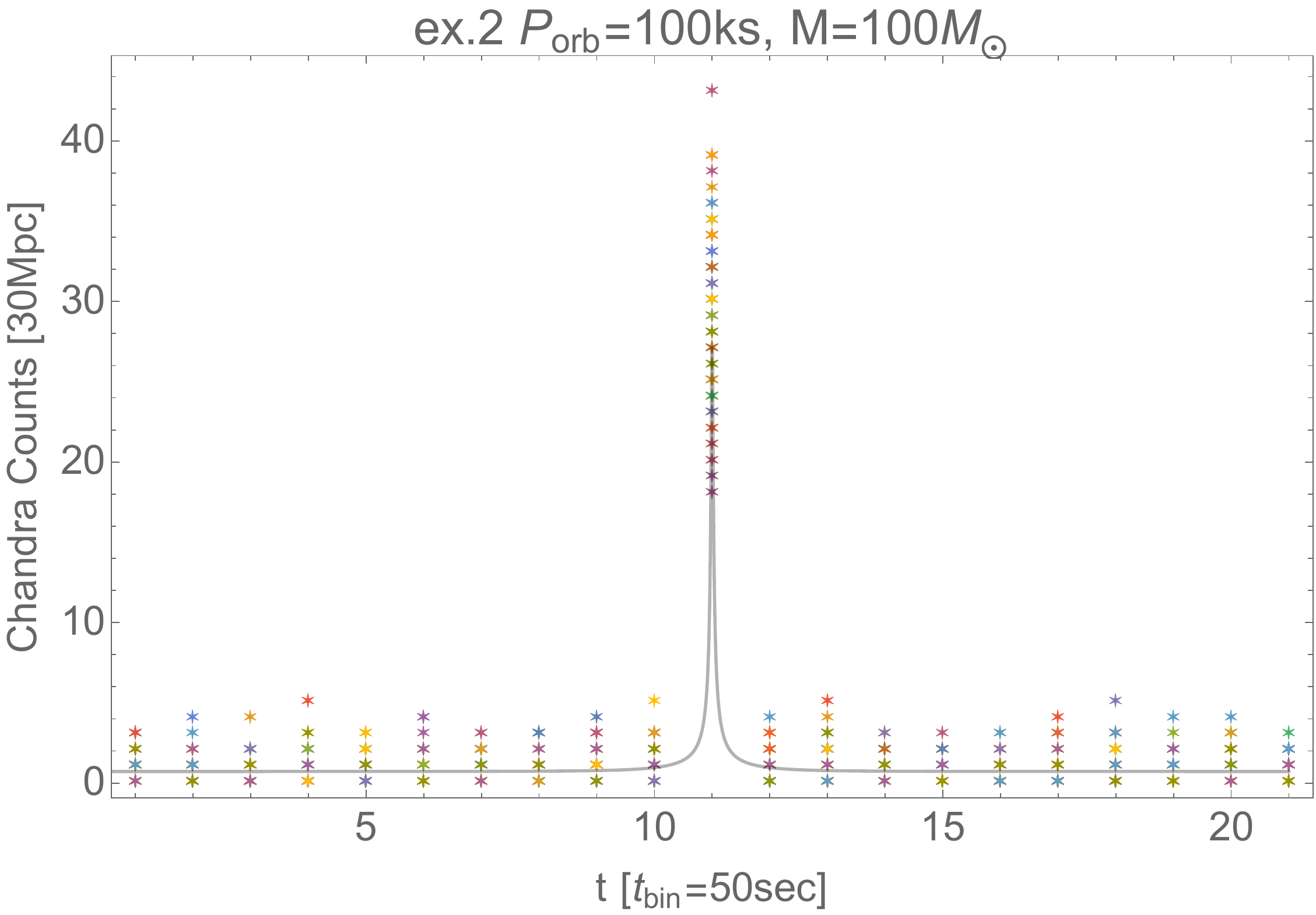} &
\includegraphics[scale=0.25]{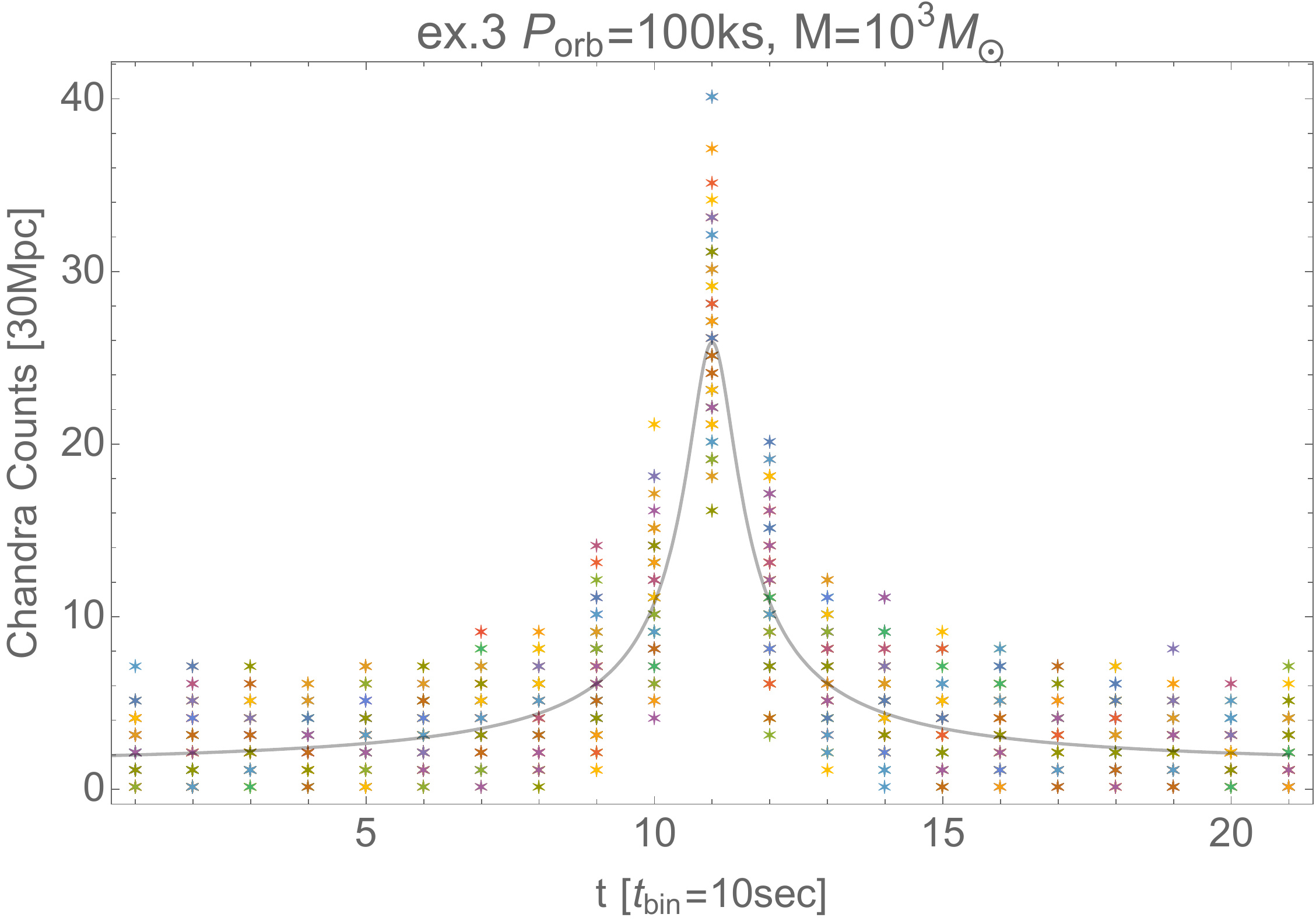} \\
\includegraphics[scale=0.25]{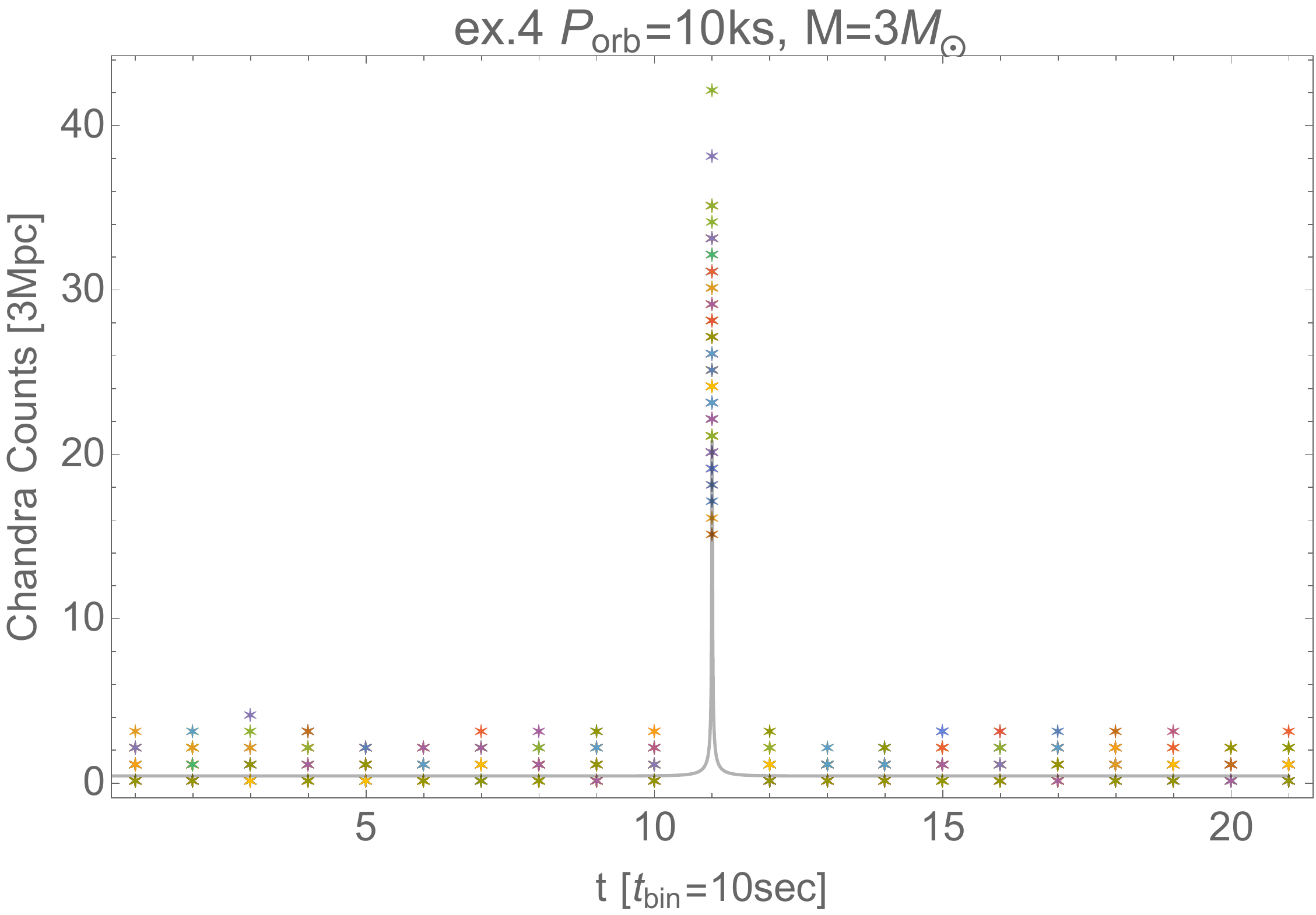} &
\includegraphics[scale=0.25]{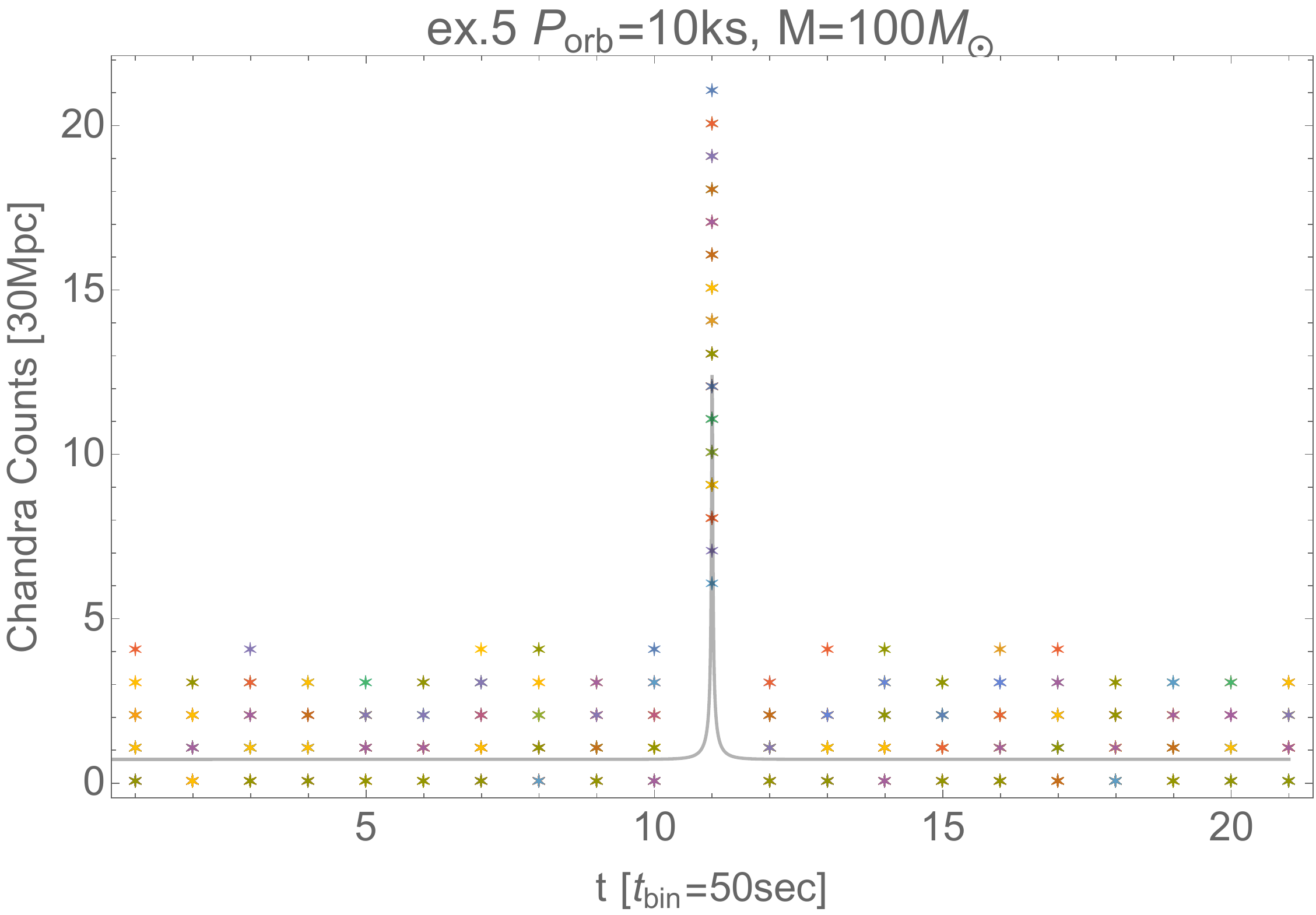} &
\includegraphics[scale=0.25]{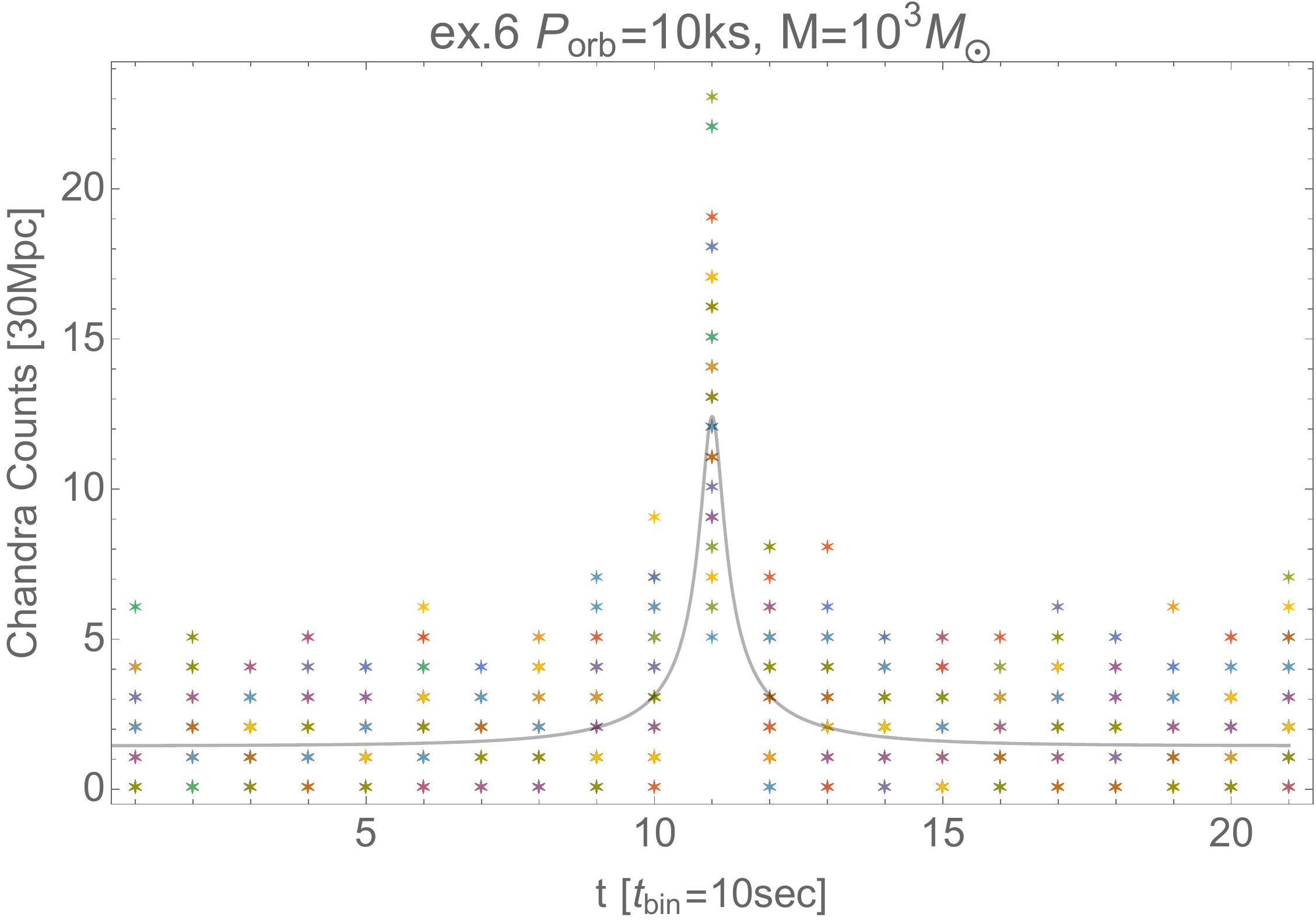}
\end{array}$
\end{center}
\vspace{0pt}
\caption{
Simulated \Chandra~light curves of six different binary systems, zoomed in on
the lensing flare. Each color represents one of the 100 realizations of the
light curve, the spread of counts at each epoch gives an approximate
uncertainty. Each panel corresponds to the binary parameters marked in the
left panel of Figure \ref{Fig:2} by green dots labeled `ex $\#$'.
}
\label{Fig:3}
\end{figure*}

Figure \ref{Fig:3} shows example light curves from the specified locations in
binary parameter space denoted in Figure \ref{Fig:2} (labeled green dots). The
axes are labeled in units of expected counts from a \textit{Chandra}-like instrument
for an Eddington luminosity source at the denoted distance, or a \Lynx-like
instrument with the same source at $10 \times$ that distance. Each panel in
Figure \ref{Fig:3} zooms in on the flare caused by lensing of an emission
region bound to one of the COs (we assume equal masses).

To compute the light curves we use the point source Doppler-boost plus lensing
models found in \S 2.3 of \citep{DoDi2018}. Denoting the Doppler+lensing magnification
model $\mathcal{M}(t)$, the expected number of counts $K$ is found from randomly
sampling the Poisson distribution,
\begin{equation}
\mathcal{P}(K) =  \frac{(t_{\rm{bin}}\mathcal{C}_0\mathcal{M}(t))^K}{K!} \exp{\left[-t_{\rm{bin}}\mathcal{C}_0\mathcal{M}(t)\right]},
\end{equation}
where $t_{\rm{bin}}$ is the bin size and
$\mathcal{C}_0$ is in units of counts per second. Then $\mathcal{P}(K)$ provides the
probability of detecting $K$ counts per $t_{\rm{bin}}$ given the average number of expected counts per bin $t_{\rm{bin}}\Cnt_0\mathcal{M}(t)$.

In Figure \ref{Fig:3}, we plot a solid grey line to represent the
magnification model, $t_{\rm{bin}}\mathcal{C}_0\mathcal{M}(t)$, and we plot
scattered points to represent the expected counts drawn from the Poisson
distribution. We realize the expected light curve 100 times (each realization
denoted by a different color) to give an indication of the uncertainty at each
observation time and, hence, the prospect of discerning the lensing flare. For
each light curve we assume an equal mass ratio binary where each binary
component has the same intrinsic luminosity. This implies that the lensing
flare occurs twice per orbital period, but also that the lensing flare is less
magnified than it would be in the case where only one of the binary components
is bright. For the same reason, other sources of emission that are not
lensed, such as that from a disk surrounding the binary, would further
decrease the magnification of the flares shown in Figure \ref{Fig:3}
\citep[see also][where the same issue is discussed in context of the orbital
Doppler boost]{PG1302Nature:2015b}.

Figure \ref{Fig:3} illustrates the distances to which lensing flares could be
identified from different types of binaries. For BNS systems within $\sim
3$~Mpc ($30$~Mpc), a \Chandra-like (\Lynx-like) X-ray telescope could identify
X-ray lensing flares, even when binning data over short, 10 second intervals
needed to temporally resolve the lens flare. For more massive binaries, the
longer flare duration and brighter intrinsic luminosity allows identification
to farther distances. A \Chandra-like (\Lynx-like) instrument can identify
lensing flares from $\approx 10^2-10^3 \Msun$ binaries out to $\sim 30$
$(300)$~Mpc.  Using longer duration time bins could allow detection out to
even farther distances than considered in Figure \ref{Fig:3}, but at the cost
of washing out the lensing flare for short orbital period systems.  We now
turn towards estimating the number of binaries within these distances, and in
the preferred range of orbital periods delineated in Figure \ref{Fig:2}.

\section{Plausibility of detection}
\label{S:Plausibility}

In this section we determine the plausibility of detecting a repeating  
self-lensing event from a CO binary by estimating the number density of CO binaries
with orbital periods $P_{\orb} \lsim 10^2$ks, and the fraction of which that are
bright enough to be detectable with present or future X-ray observations. Non-
repeating systems may be even more prevalent, though less certain in their
identification, as discussed previously (see the discussion of Figures
\ref{Fig:1} and \ref{Fig:2}).

While the rate of CO mergers is beginning to be pinned down by LIGO/VIRGO, it
is still orders of magnitude uncertain (10's to 1000's Gpc$^{-3}$ yr$^{-1}$.
Furthermore, the inference from merger rate to the number density of systems
at larger separations, where lensing signatures manifest, depends on the
poorly understood rate at which CO-binary orbits decay and, hence, the
mechanisms that bring CO binaries close enough to merge within a Hubble time.
These mechanisms also dictate the length of time that the binary can be bright
enough for observable lensing.  
While we do not tackle rate estimates in
detail here, we demonstrate that such systems could be waiting to be
discovered in the local universe. Whether or not any such systems exist rests
primarily on the unknown fraction of CO binaries that do not merge within a
Hubble time and the fraction of time that a CO binary emits near its Eddington
luminosity. Null detections could help to narrow such uncertainties. We end
this section by briefly considering mechanisms that
can generate bright EM emission.

\subsection{Rates}

We proceed by estimating the number of CO binaries per orbital period and
binary mass. We multiply this by the probability for lensing (Eq.
\ref{Eq:Prob1}) and the fraction of CO binaries that are EM bright and then
integrate over all relevant binary masses and observationally accessible
orbital periods. We consider two illustrative cases:
\begin{enumerate}
	\item Isolated CO binaries driven to merger solely by GW decay.
	\item Dynamically formed binary black holes (BBHs) that merge within globular clusters.
\end{enumerate}

\subsubsection{Isolated GW-driven binaries}
\label{S:Rates:Iso}

For the isolated CO binaries, we first estimate $d n_{\rm gal}/dP_{\orb}$, the
number of binaries per Milky-Way-like galaxy with orbital periods between
$P_{\orb}$ and $P_{\orb}+dP_{\orb}$. As this quantity is very
uncertain, we decide, within the context of this work, to take a simplest
approach to its estimation. Instead of appealing to complex population
synthesis calculations, which take as input many uncertain analytical
prescriptions for uncertain orbital decay mechanisms, we assume that the CO
binaries are on circular orbits and that only GW emission acts to change the
binary orbital elements. We further assume that the population is in steady
state. We recognize that each of these assumptions may break down for the
systems that we consider, however, this simplest case serves as a
baseline, with which we compare results from a second estimate based on a
specific formation channel in the next subsection. With these assumptions,
we solve the homogeneous advection equation that describes the evolution of
CO- binary orbital periods over time \citep[see also][]{ChristianLoeb:2017},
\begin{equation}
\partial_{P_{\orb}} \left[\left.\frac{dP_{\orb}}{dt}\right|_{\rm GW} \frac{d n_{\gal}}{d P_{\orb} } \right] = 0.
\end{equation}
This has solutions,
\begin{eqnarray}
 \frac{d n_{\gal}}{d P_{\orb} } &=& \frac{C}{A} \frac{P^{5/3}_{\orb}}{M^{5/3}} \\ \nonumber
 A &\equiv& \frac{384}{5} (2 \pi)^{8/3} \frac{G^{5/3}}{c^5} q_s,
\end{eqnarray}
where $C$ is a constant of integration that we set via the LIGO-rate boundary condition.

We incorporate a distribution of binary masses by assuming that each
total binary mass is drawn from a Salpeter mass function of the form $dN/dM
\propto M^{-\alpha}$. Then we define,
\begin{eqnarray}
p(M) \equiv \frac{1}{N_{\rm tot}}\frac{dN}{dM} = 
 \frac{1 -\alpha}{M^{1-\alpha}_{\max} - M^{1-\alpha}_{\min}} M^{-\alpha} ,
\end{eqnarray}
where $N_{\rm tot} = \int^{M_{\max}}_{M_{\min}}{\frac{dN}{dM} \ dM}$.

To fix the constant of integration, we impose a boundary condition requiring
the flux of BBHs passing into the LIGO band be equal to the LIGO co-moving
merger rate. Assuming that the constant $C$ is mass independent (that the mass
dependent  co-moving merger rate is flat between $M_{\min}$ and $M_{\max}$),
and because $p(m)$ integrates to $1$, we find that $C = \mathcal{R}_i$. Where
the co-moving merger rate, $\mathcal{R}_i \equiv dn_i/dt$, is estimated by
LIGO for the i$^{\rm th}$ type of CO binary ($i\rightarrow$ BBH, BH-NS, 
BNS; see Table \ref{Table:rates}). This rate density is then converted into
a rate per Milky-Way-like galaxy by assuming that there are $dN_G/dV = 10^{-2}$
galaxies per Mpc$^{3}$ \citep{MDPRada:2009}. Our solution for the number of CO
binaries per galaxy, orbital period, and mass is then,
\begin{eqnarray}
\frac{d^2 n_{\gal}}{dP_{\orb}dM} &=& \mathcal{R}_i  \frac{p(M)}{A}  \left( \frac{P_{\orb}}{M} \right)^{5/3}.
\end{eqnarray}

Next we multiply our solution by the mass and orbital period dependent
probability of a significant lensing event, $\mathcal{D}_{\obs}
\mathcal{P}_L(P_{\orb},M)$,  where the second term is the probability of a
lensing flare given that an entire orbit is observed (Eq.~(\ref{Eq:Prob1}))
and $\mathcal{D}_{\obs}$ is the fraction of an orbit that is observed with a cadence higher than once per flare duration. We then
integrate over all binary masses and the range of orbital periods that are of
observational interest, and multiply by the total number of galaxies per
volume,
\begin{widetext}
\begin{eqnarray}
\Num_{\rm{lens}}(P_{\max}) = \mathcal{D}_{\obs} f_{\rm EM} \Delta \Omega_{P_{\max}} \frac{dN_{G}}{dV} 
\int^{P_{\max}}_{0}{
	\int^{M_{\max}}_{M_{\min}}{  
		\int^{z_{\max}(P_{\orb},M)}_{0}
 			\frac{d^2V}{dz d\Omega} dz \frac{d^2 n_{\gal}}{dP_{\orb}dM} \mathcal{P}_L(P_{\orb},M) \ dM} \ dP_{\orb}}, 
 \label{Eq:Nlens}
\end{eqnarray}
\end{widetext}
where we have introduced the parameters $\chi$, the fraction of all CO
binaries that will merge; $\Delta \Omega_{P_{\max}}$, the area of sky observed for at
least the duration of $P_{\max}$; and $f_{\rm EM}$, the fraction of a binary
lifetime (for orbital periods shorter than $P_{\max}$) that the binary is EM
bright, where in this case bright means emitting near the Eddington
luminosity. Here $z_{\max}$ is the maximum redshift out to which one could
observe the EM emission. This is found by solving for a maximum luminosity
distance by setting the minimally magnified count-rate (Eq. (\ref{Eq:Bkg_cnt})
multiplied by a minimum magnification factor of $1.34$), equal to a minimum
detectable count rate given by the condition that $\mathcal{C}\tau_{ev} = 1$.
Hence, $z_{\max}$ is a function of $M$, $P$, and $q$ and represents the
maximum redshift where at least one count can be collected during the
(minimally) lensed flare. The  quantity $d^2V/(dzd\Omega)$ is the co-moving
volume per redshift and solid angle.

For BBHs and BNSs we assume a mass ratio of unity, choosing $M_{\min}=10 \Msun$
and $M_{\max}=100 \Msun$ for BBHs, and choosing $M_{\min}=2 \Msun$ and $M_{\max}=6
\Msun$ for BNSs. For BH-NS binaries, we allow the total binary mass to range from
$M_{\min}=5 \Msun$ and $M_{\max}=50$, and then assign a mass ratio by assuming
a $1.4 \Msun$ NS.

The results of computing the integral in Eq. (\ref{Eq:Nlens}) are listed in
Table \ref{Table:rates} for both  \Chandra-like and \Lynx-like sensitivities,
assuming our fiducial value of $P_{\max}=100$~ks. Assuming small $d_{\max}$
($\lsim 100$'s Mpc), the results in Table \ref{Table:rates} scale as,
\begin{widetext}
\begin{eqnarray}
\Num_{\rm{lens}} = 
    \begin{cases} 
     1 \times 10^4    &(\rm{BBH})  \\
     2 \times 10^4    &(\rm{BH-NS})  \\
     3 \times 10^3    &(\rm{BNS}) 
     \end{cases} 
     \times \frac{\Delta \Omega_{100}}{4 \pi} \frac{\mathcal{D}_{\obs} f_{\rm EM}}{\chi} \left(\frac{\mathcal{R}}{10^{-5} \rm{gal}^{-1} \rm{yr}^{-1} }\right) \left(\frac{F_{\min}}{F_{\min, \Chandra}}\right)^{3/2} \left(\frac{10^2 \rm{ks}}{P_{\max}}\right)^{10/3} ,
\label{Eq:NlensScale}
\end{eqnarray}
\end{widetext}
where we have defined a minimum flux sensitivity $F_{\min}$ compared to the
sensitivity of \Chandra. For \Lynx, which we take to be $100 \times$
more sensitive than \Chandra, these numbers increase by a factor of $10^3$. 

The number of possible lensing binaries increases steeply with the maximum
allowed binary orbital period. The $P_{\max}$ dependence can be determined
from Eq. (\ref{Eq:Nlens}): because $\mathcal{N}_{\rm{lens}} \propto
(dn^2/dP_{\orb}dM ) (\mathcal{P}_L) (d^3_{\max}) dP_{\orb} \propto
P^{5/3}_{\orb} P^{-1/3}_{\orb} P^1_{\orb} P^1_{\orb} = P^{10/3}_{\orb}$. That
is, the number of lensing candidates increases with $P$ because. a) there are
more binaries at longer orbital periods because they spend a longer time there
and b) because longer orbital periods imply longer lensing durations which
allow a larger detection volume $\sim d^3_{\max}$ through the requirement that
$\mathcal{C}\tau_{\rm{ev}} \geq1 $. This increase in number with $P_{\max}$
outweighs the weaker decrease in lensing probability with $P_{\max}$.

This scaling holds only as long as continuous observations can be made for
duration $P_{\max}$. For orbital periods much longer than that of a typical
observation time, the factor $\mathcal{D}_{\obs}$ is less than one, and the
full $P^{10/3}_{\max}$ increase is not realized. Because $\mathcal{D}_{\obs}
\propto \min\left[T_{\rm obs}/P, 1\right]$ for total observation time $T_{\rm
obs}$, the scaling goes as $P^{7/3}_{\max} T_{\rm obs}$ for $P_{\orb}>T_{\rm
obs}$. Additionally an instrument specific drop in number with $P_{\max}$ will
occur if $\Delta \Omega_{P_{\max}}$ decreases steeply with $P_{\max}$ above
some value.

To further understand the numbers predicted by Eq. (\ref{Eq:Nlens}) and
recorded in Table \ref{Table:rates}, we consider, as an example, the BNS
population. At $P_{\orb}\leq10^2$~ks, our steady-state model, tied to a LIGO
rate of $10^3$Gpc$^{-3}\rm{yr}^{-1}$ ($10^{-4}$ gal$^{-1}$yr$^{-1}$), implies that
there are $\sim10^6$ BNSs per galaxy. Because the merger time is
approximately a Hubble time for a BNS with $P_{\orb} \rightarrow 10^2$~ks,
this means that a galaxy harbors $\sim10^6$ BNSs over its lifetime. The
lensing probability reduces this number by a factor of $\sim 600$, leaving a
few $\times 10^3$ near-line-of-sight aligned BNSs per galaxy. At a distance of
$\sim7$~Mpc for \Chandra~($\sim70$~Mpc for \Lynx) there are $\sim10$
($\sim10^4$) galaxies. Multiplying by $10^3$ lensing candidates per galaxy
yields the (all-sky) numbers for BNS lensing candidates in Table
\ref{Table:rates}.

Because such binaries could be part of an observable, high-mass X-ray binary
population before they become BNSs, it serves as a consistency check to
consider the implications of our rate calculation for the number of high-mass
X-ray binaries expected per galaxy. If $10^6$ BNSs are formed per galaxy per
Hubble time, and if winds and mass overflow can produce X-ray emission via
accretion onto the first-formed CO for $10^4-10^5$~yr, then we expect 1-10
high-mass X-ray binaries per galaxy, which is not unreasonable
\citep[\textit{e.g.}][]{Fabbiano:2019}.

Eq. (\ref{Eq:NlensScale}) and an estimate for $\Delta \Omega_{100}$, can be
used to constrain the quantity $f_{\EM}/\chi$ even via the non-detection of
lensing flares. This quantity elucidates the number of Eddington luminosity CO
binaries with orbital periods below $100$~ks, and also the fraction of such CO
binaries that merge vs. stall. This is interesting because the orbital period
of a circular, equal-mass-ratio binary that will merge in a Hubble time is $P
\leq 410 [M/(60\Msun)]^{5/8}$ks. For \Chandra, $\Delta \Omega_{100}/(4
\pi)\sim 0.01$ \citep{CSC:2010}, hence, we would expect to see at least one
lensing systems with \Chandra~if $f_{\EM}/\chi\gsim10^{-2}$, and assuming a
similar $\Delta \Omega_{100}$ for \Lynx, if $f_{\EM}/\chi\gsim10^{-5}$. We
offer estimates for $f_{\EM}$ in \S \ref{S:EMmechs}.

\subsubsection{Dynamically formed BBHs}
\label{S:Rates:Dyn}

We next narrow our discussion from all CO binaries to only BBHs, and specifically those that
form dynamically in globular clusters (GCs). The dynamical formation channel
predicts a specific number of BBHs per GC and also the period distribution of these BBHs.
\cite{SamsingTDE+2019} show that
\begin{eqnarray}
\left. \frac{dn_{\gal}}{d  P_{\orb} dM} \right|_{\rm dyn} &\propto& \frac{ p(M) }{P^{5/3}_{\orb}} \exp{\left[-\xi_{\rm{HB}}  \frac{1 - \epsilon \left(P_{\orb}/P_{\rm{HB}}\right)^{-20/21} }{1 - \epsilon}\right]}  \nonumber \\ 
\xi_{\rm{HB}} &=& \left( \frac{8192}{3645 \pi} \frac{v^{11}_d}{c^5(GM)^3 n_{\rm GC}} \right)^{2/7} \nonumber \\ 
P_{\rm HB } &\approx& \frac{3\sqrt{3} \pi }{4} \frac{G M }{v^3_d} \qquad \epsilon \equiv \left(\frac{7}{9}\right)^{-10/7}
\label{Eq:dPGC}
\end{eqnarray}
which introduces the GC velocity dispersion $v_d$ and stellar number density
$n_{\rm GC}$ for which we take fiducial values of $v_d=12.5 \rm{km}s^{-1}$ and
$n_{\rm GC}=10^5$~pc$^{-3}$ \citep[see, \textit{e.g.}, ][and references
therein]{SamsingTDE+2019}.  Here we have assumed that binaries are all equal
mass due to the propensity for mass segregation and exchange interactions to
produce equal mass ratio binaries \citep[\textit{e.g.}][]{HeggieHut:2003}.

With Eq. (\ref{Eq:dPGC}) we repeat the calculation of \S \ref{S:Rates:Iso} for
GW-driven, isolated binaries, but without using the LIGO rate to tie our
estimate to only those systems that will merge within a Hubble time. To do so,
we assume that all GCs will have approximately $5$ BBHs in their core at any time
\citep[\textit{e.g.},][]{KremerGCs+2019} and that every Galaxy has 150 GCs
\citep{Harris_MWGCs:2010}. Then we normalize the number of dynamically formed BBHs
per orbital period, $dn_{\gal}/d\log P |_{\rm dyn}$, by setting $5 \times 150$
equal to the integral of the RHS of Eq. (\ref{Eq:dPGC}) over all orbital
periods up to the hard binary limit, $P_{\rm HB}$.

\begin{figure*}
\begin{center}
\includegraphics[scale=0.35]{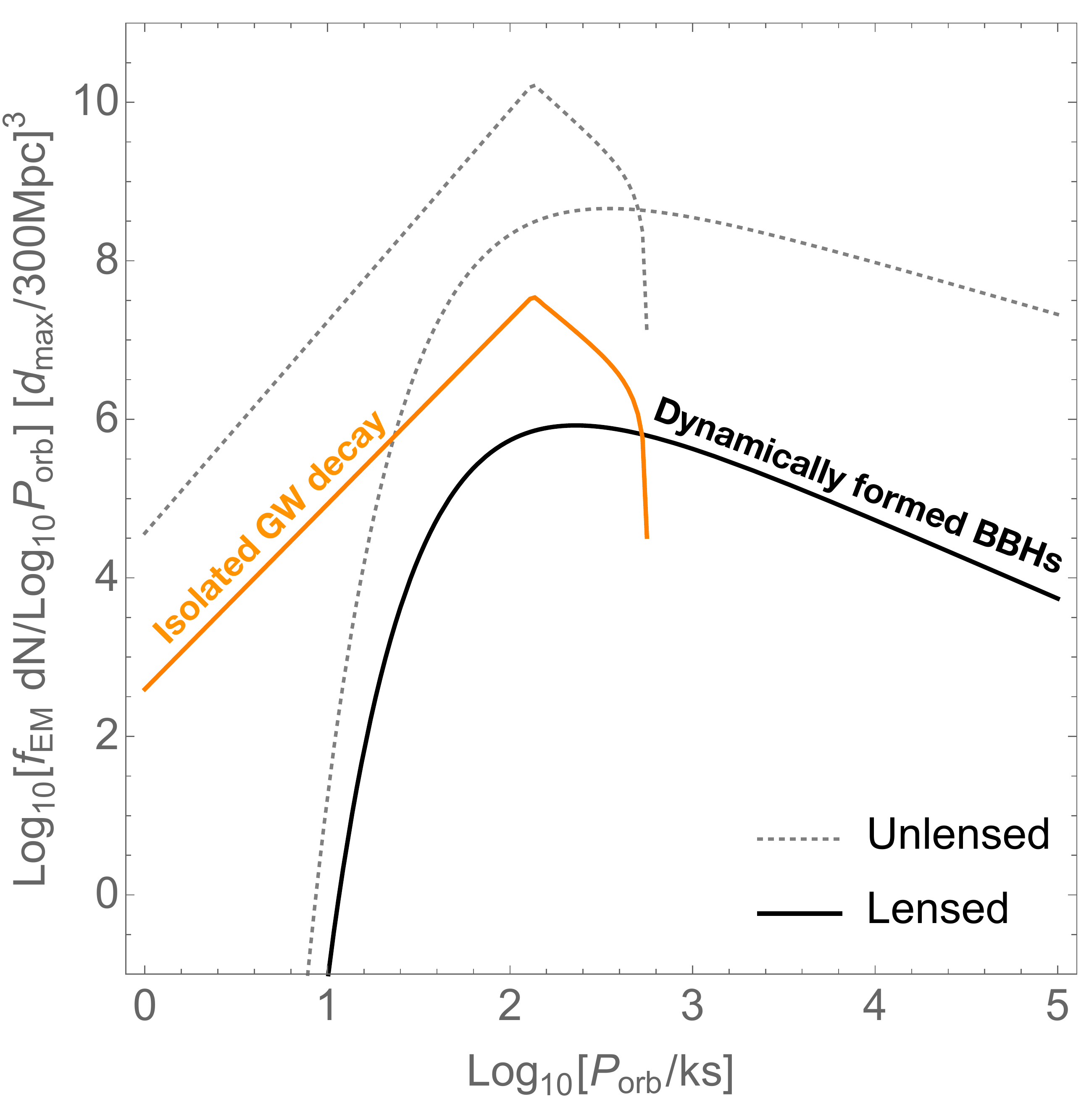} 
\end{center}
\vspace{0pt}
\caption{
The number of compact-object binaries per log orbital period integrated over a
range of binary masses. The orange lines are drawn for a population of compact
objects undergoing orbital decay due only to GW radiation (\S
\ref{S:Rates:Iso}) and are truncated at the maximum orbital periods for which
GW decay could merge the pair in a Hubble time (assuming a binary mass ratio
of $q=1$). The black lines represent the dynamically formed population of BBHs
described in \S \ref{S:Rates:Dyn}. The gray dotted lines are drawn without
taking into account the probability for orbital alignment to cause strong
lensing. The probability for a binary to be EM bright is set to $f_{\EM}=1$
for scaling purposes, in reality this number might be $10^{-5}$, but is
uncertain.
}
\label{Fig:4}
\end{figure*}

The period distributions for self-lensing BBH systems, formed via both
dynamical and isolated channels are plotted in Figure \ref{Fig:4} for a range
of binary masses. The orange lines are drawn for isolated GW decay matched to
a LIGO rate of $100$ Gpc$^{-3}$ yr$^{-1}$. The black lines are drawn for the
dynamically formed BBHs. The dotted-gray lines illustrate the period distribution
before taking the lensing probability into account.

The dynamically formed BBH population peaks in number at orbital periods of a
few 100~ks, ideal for the X-ray self-lensing observations, but has a lower
total number than the LIGO normalized population, consistent with the low end
of the LIGO BBH rates (See Table \ref{Table:rates}).

\begin{table*}
\begin{tabular}{l|l|l|c|c|c|c}
   \multicolumn{6}{c}{  \quad $P_{\max} = 10^2 \rm{ks}$  \quad $\mathcal{D}_{\obs}=1.0$}
  \\
  \hline
  Binary type  
  & $\mathcal{R}_i \ $  [Gpc$^{-3}$ yr$^{-1}$]
  & \multicolumn{2}{c}{---Number of \textbf{\textit{Lynx}} Lensing Candidates---} & \multicolumn{2}{c}{---Number of \textbf{\textit{Chandra}} Lensing Candidates---} \\
&
  & Isolated  $\left[\frac{\Delta \Omega_{100}}{4 \pi} f_{\EM}/\chi \right]$
  & Dynamical $\left[\frac{\Delta \Omega_{100}}{4 \pi} f_{\EM} \right]$
  & Isolated  $\left[\frac{\Delta \Omega_{100}}{4 \pi} f_{\EM}/\chi \right]$
  & Dynamical $\left[\frac{\Delta \Omega_{100}}{4 \pi} f_{\EM} \right]$
   \\
\hline
  BBH & $10-100$     & $(10^6-10^7)$  & $\sim10^5$  &   $(10^3-10^4)$&  $\sim10^2$ \\
  BH-NS & $\leq600$   & $ \leq 10^8 $ & -- &  $ \leq  10^5 $ & -- \\
  %
    BNS & $100-3800$  & $ (10^6-10^8) $ & -- & $ (10^3- 10^5) $  &-- \\
\end{tabular}
 %
 %
\caption{
The approximate, all-sky number $\Num_{\rm{lens}}$ of lensing systems by
binary type and formation mechanism multiplied by the fraction $f_{\EM}$ that
emit at or above the Eddington luminosity and by the solid angle $\Delta
\Omega_{100}$ of sky observed for at least the duration of the maximum orbital
period set to $P_{\max} = 100 \rm{ks}$. The quantity $\chi$ is the fraction
that will merge within a Hubble time and is only relevant for the "Isolated" category which is tied to the LIGO merger rate. Merger rate ranges are from \citet{LIGO_O2_COBs:2018}.
}
\label{Table:rates}
\end{table*}

\subsection{EM emission mechanisms}
\label{S:EMmechs}

While there are a number of known X-ray bright accreting compact objects,
namely the X-ray binaries \citep[\textit{e.g.},][]{CharlesCoe:XRBins2003}, the
only known X-ray bright double compact-object binaries are double neutron star
systems identified by the existence of a pulsar in the binary
\citep[\textit{e.g.},][]{Burgay+2003, ChaterjeeXrayDNS+2007}.  As the rate
estimate of the previous sub-sections shows, this lack of EM identification of
double compact-object binaries may be due to scarcity within the presently
observable volume. It is also possible that without a unique identifier of the
binary (as exists for the double neutron star system containing a pulsar),
double compact-object systems may be confused with single compact-object
systems, the point of our work here being to demonstrate self-lensing flares
as possible identifiers. A final complication could be the duration for which
systems at detectable orbital periods are EM bright, which we discuss
presently.

The opportunity to use self-lensing as a tool to study CO binaries exists only
if one or both COs emit bright EM radiation, a property we parameterized by
the EM duty cycle $f_{\EM}$ in the previous subsections. One can estimate the
EM duty cycle with the simple approximation $f_{\EM}\sim\tau_{\EM}/\tau_{\rm
merge}$, which relies on an estimate of how long any member of a CO binary can
be bright enough for detection and so depends on the distance from the observer
and the binary properties, as well as the EM-radiation-generating mechanism
itself. Because this is uncertain, and indeed something that an ongoing search
for self-lensing flares could constrain, here we simply motivate a few ways
that CO binaries could emit near their Eddington luminosities.

While NSs and magnetars can be bright after birth due to magnetically driven
spin down power or fallback accretion, they typically only reach sub-Eddington
luminosities that render them detectable within our own Milky Way and its
satellites. \citep{KaspiBelo:2017, Chatt_FBD_AXP:2000}. One exception is the
Ultra Luminous X-ray sources that could be super Eddington accretors
\citep{ULXrev:2017}. However, as the physical nature of these objects is
uncertain, we do not attempt to predict their prevalence in binaries or their
EM duty cycle here.

BBHs can also be bright for limited
times due to fallback accretion \citep{PernaDUff+2014}. 
A fallback disc could contain $M_d \equiv q_d M$ of material, where M is the
total binary mass. If the disc can accrete onto the binary at a fraction
$f_{\Edd}$ of the Eddington rate, $\dot{M} = f_{\rm{Edd}} 2.3 \times 10^{-8}
\Msun \rm{yr}^{-1} (M/\Msun)$, assuming $10\%$ accretion efficiency, then the
disc could have a total active lifetime
\begin{eqnarray}
\tau_{\EM} &\approx& \frac{q_d M}{f_{\Edd} 2.3 \times 10^{-8}
\Msun \rm{yr}^{-1} (M/\Msun)} \\ \nonumber
&\approx&  4.3 \times 10^{5} \rm{yr} \ f^{-1}_{\Edd} \frac{q_d}{10^{-2} },
\end{eqnarray}
which is the fraction $q_d$ of the mass doubling time at rate $f_{\Edd}$. From this
we estimate a duty cycle for EM bright emission $f_{\EM} \lsim 10^{-5}$,
assuming that a binary merges in a Hubble time. We note that \citet{Perna_fallbackBBH:2016}
and \citet{Kimura_fallbackBBH:2017} explore the possibility of dormant
fallback discs being reborn by tidal torques in BBH systems. Specifically, \citet{Kimura_fallbackBBH:2017} find that fallback discs around each BH can
be revitalized $\sim10^5$ years before merger, but only with the brightest,
super-Eddington stage of accretion happening less than a year before merger.

The formation of the second CO may also occur soon after a 
common-envelope phase, supplying gas to the CO binary \citep[see][for a
review]{Ivanova+2013}. It may be possible for residual accretion to continue
to produce EM radiation after the common envelope is ejected.

Another intriguing possibility is accretion onto the CO binary from a
stellar companion in an hierarchical triple system. It is increasingly clear that
massive stars are born in binaries, and also in hierarchical triples
\citep[\textit{e.g.},][]{MoeDi:2017}. The existence of a third star may play
an important evolutionary role in allowing the compact binary to form. For
example, several studies have invoked secular dynamical interactions (particularly the
Kozai-Lidov mechanism) with a wide-orbit companion to shrink the inner orbit
so that merger within a Hubble time is possible.

As the third star in an un-evolved star plus CO-binary triple evolves, it will
emit winds. In some cases, it may come to fill its Roche lobe. In either case,
a fraction of the mass from the outer star will come under the gravitational
influence of the stellar remnants in the inner binary. As they fall toward the
compact remnants, they can become X-ray bright. The duration of mass transfer
is determined by the evolutionary time scale of the outer star. For stars of
high mass, one or both components of the inner binary may be X-ray bright for
$\sim 10^5$~yr, while the time scale is longer, up to $\sim 10^7$~yr, for
stars of lower mass. An interesting point is that the time to merger evolves
as mass is transferred. Changes by more than two orders of magnitude may not
be unusual \citep{RDS:2018}. Thus, typical values of the duty cycle $f_{\EM}$
may be as large as a few percent to unity.

Recently, \citet{SamsingTDE+2019}, \citet{LopezTDEsim+2018}, and
\citet{KremerTDE+2019} pointed out that in the dynamical formation scenario,
BBHs could interact with stars or even planets \citep{KremerGCPlanets+2019},
allowing both or one BH to become bright via tidal disruptions. Such events could
exceed the Eddington luminosity, though for short durations.

While here we have simply sketched a motivation for the possibility of bright EM
emission from CO binaries, future work can be carried out to better quantify
the possibility for bright (Eddington) and long lived ($\gsim10^{-5}$ of an
average merging binary's lifetime) EM emission caused by a combination of the
above possibilities.

\section{Discussion}
\label{S:Discussion}

We have shown that binary self-lensing of a bright component of a CO binary by
the other could serve as a unique identifier of the progenitors of LIGO and
LISA GW sources, namely the stellar-mass-CO inspirals and mergers. We estimate
the timescales, magnification, and possible rates of such events. Binaries
with periods of order $100$ ks can exhibit repeating self-lensing flares with
maximum magnifications ranging from a few to $\times100$, depending on binary
mass. They are detectable within single X-ray observation times, and can merge
via GW emission within a Hubble time. Longer orbital period systems (with
lensing properties given by systems to the right of the vertical lines in
Figure \ref{Fig:1}) may be more abundant, though detectable as single-flare
events that can be followed-up to search for the predicted succeeding flares.

We estimate that, out to a distance to which a \Lynx-like future X-ray
observatory could see an Eddington limited source, $\sim10^7f_{\EM}/\chi$ CO
binaries, across the entire sky, would have orbital periods of $\lsim100$~ks
and be inclined sufficiently to the line of sight to cause strong lensing
flares. The same estimate using the flux-sensitivity of \Chandra~is
$\sim10^4f_{\EM}/\chi$. Here $f_{\EM}$ parameterizes the unknown duty cycle of
CO binaries below a maximum orbital period that have at least one bright
component (bolometric luminosity at the Eddington luminosity or greater), and
$\chi$ is the fraction of CO binaries at of order $100$~ks orbital periods
that will merge rather than stall. For example, $\chi$ could be less than one
if gas dynamics slow or prevent merger \citep{Tang+2017, Derdzinski+2018,
MunozMirandaLai:2019, MoodyStone+2019}, or if the binary is broken up during
dynamical formation \citep[as happens for the 2-body and 3-body mergers
of][]{SDII:2018, DSIII:2018}.

While the above estimates invoke a steady-state population of GW-driven CO
binaries, we also considered a population of BBHs that are driven to merger in
GCs via dynamical processes. The total number of lensing candidates predicted
via the dynamical channel is consistent with the low-end predictions of the
"Isolated" GW-driven channel. It is interesting to note that the two
populations have significantly different period distributions that, if probed
by a self-lensing population, could differentiate between the two formation
scenarios \citep[see also][]{SamsingTDE+2019}.

Furthermore, within a given CO-binary formation model, we can constrain the value of
$f_{\EM}/\chi$ by searching for such repeating flares in present X-ray data.
Considering the isolated GW-decay model, null detection with  
\Chandra-sensitivity archival data covering $1\%$ of the sky would constrain this
quantity to be $f_{\EM}/\chi\lsim 10^{-2}$ and null detections with a similar
sky coverage at \Lynx-sensitivity would require $f_{\EM}/\chi \lsim 10^{-5}$
(see Eq. \ref{Eq:NlensScale}), which begins to become astrophysically interesting
given expectations for certain astrophysical scenarios
discussed in \S \ref{S:EMmechs}.  A wide area X-ray survey with similar
sensitivities would improve these constraints by up to two orders of
magnitude. If independent measurements of the merging fraction $\chi$ can be
made via, \textit{e.g.}, LISA vs. LIGO population analysis, then the EM duty
cycle of CO binaries can be constrained.

More interesting would be an observation of a self-lensing event. From such an
observation we can learn the orbital period and mass of the binary and its
components. For multiple flare detections in a single system, precise
phasing of the orbit allowed by flare detection could allow the
determination of orbital precession caused by, \textit{e.g.}, surrounding matter
or relativistic effects. Identification of multiple binaries via self-lensing would
allow us to infer the orbital period distribution of CO binaries at large
separations and compare to models such as those plotted in Figure \ref{Fig:4}.
This would open a window into the CO-binary population at a time in their
evolution before any GW instrument could detect them. It could also probe the
population near formation; this is because $\mathcal{O}(100)$~ks orbital
periods, which are ideal for   self-lensing detection, are also approximately
the orbital periods that delineate merging systems from those that do not
merge via GW emission within a Hubble time. For example, equal mass ratio,
$3\Msun$ BNSs on circular orbits will merger in a Hubble time when starting
from $P_{\orb}\sim60$~ks while $60\Msun$ BBHs will merge in a Hubble time
starting from $P_{\orb}\sim400$~ks. This suggests that observing BNSs near
$100$~ks orbital periods can teach us something about the fraction of BNSs
that merge (our parameter $\chi$).

While we have considered specifically \Chandra- and \Lynx-like instruments in order to
observationally ground this study, it is important to consider the interplay
between the sky coverage as a function of observation time
(setting $\Delta \Omega_{P_{\max}}$), and the instrument sensitivity (setting
$d_{\max}$). Reading off the scaling of sky coverage and sensitivity
from Eq. (\ref{Eq:NlensScale}), we can apply our calculation to other X-ray
instruments.
Compared to \Chandra~we find,
\begin{align}
\Num_{\rm{lens}, \Lynx}  &\approx 10^3 \Num_{\rm{lens}, \Chandra} \nonumber \\
\Num_{\rm{lens}, \Athena}  &\approx 30 \Num_{\rm{lens}, \Chandra} \nonumber \\
\Num_{\rm{lens}, \Axis}  &\approx 30 \Num_{\rm{lens}, \Chandra}
\end{align}
where, as throughout the rest of this work we have estimated a $100\times$
better sensitivity with \Lynx. We have estimated a $10\times$ better
sensitivity for \Axis and \Athena \citep{ATHENA:2013, AXIS:2017}, with similar
end-mission values of $\Delta \Omega_{100}$ for both (while the \Athena
mission has a larger field of view than \Chandra, its mission duration may be
shorter). While \textit{eRosita} is an all-sky survey, it will only observe
$\sim 0.01$ percent of the sky for greater than $10$~ks, and with a similar
sensitivity to \Chandra~\citep{eRosita:2012}. We also estimate that XMM Newton
and SWIFT perform comparably to \Chandra~for such a task. Ultimately, the
combination of data from all of these instruments will allow the most
constraining power.


Systems that exhibit periodic self-lensing will also exhibit periodic
modulations due to the relativistic Doppler boost. Because, we are considering
systems with orbital periods $\gsim 1$~ks, and with mass below $\sim10^3
\Msun$, magnification due to the Doppler boost never becomes as large as in
the lensing case. However, the range of binary inclination angles for which
moderate Doppler-boost magnification occurs is wider \citep[][]{DoDi2018}, and
so the probability of finding a moderately amplified, periodic, Doppler-
boosted system can be higher.

\begin{figure}
\begin{center}
\includegraphics[scale=0.23]{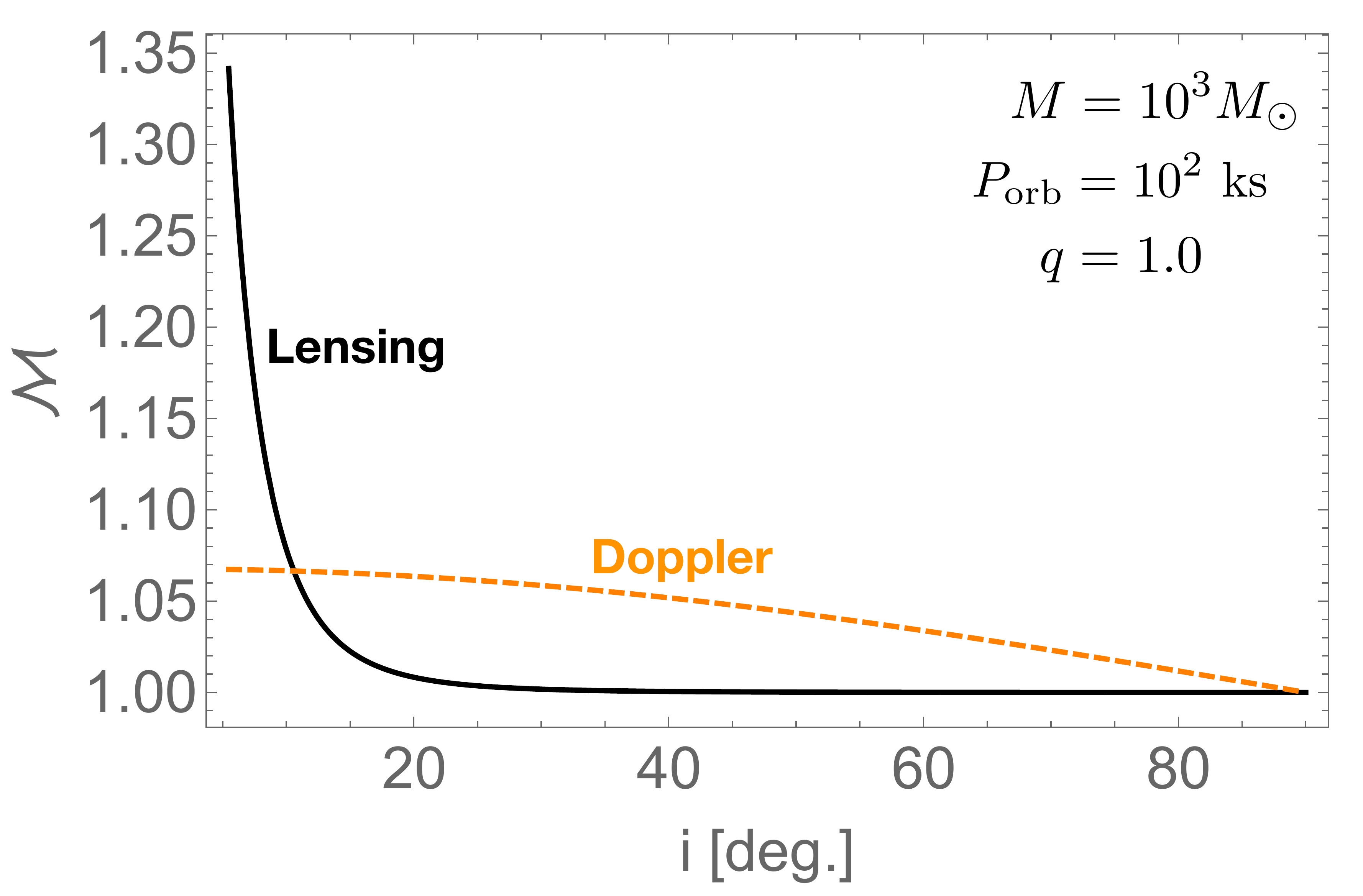}
\end{center}
\vspace{0pt}
\caption{
The magnification due to self-lensing (black-solid line) and the relativistic
Doppler boost (orange-dashed line) as a function of inclination of the binary
to the line of sight. The smallest inclination angle is chosen in order to
align the source and the Einstein radius of the lens (strong lensing limit),
smaller inclination angles result in lensing amplifications upwards of $100$ for the smallest source sizes (see Figure \ref{Fig:2}). }
\label{Fig:DopInc}
\end{figure}

As an example, consider the approximate amplitude of Doppler modulations,
$(3-\alpha)v_{\orb}/c \cos{i}$ for line-sight inclination $i$ and spectral
slope $\alpha$. We choose $\alpha=1$ for demonstration. For a $M=10^3 \Msun$
binary, for which Doppler amplification would be the largest, Figure
\ref{Fig:DopInc} plots the line-of-sight inclination angle dependence of the 
Doppler-boost magnification as well as lensing magnification for
the same binary system. From Figure \ref{Fig:DopInc}, it clear that lensing
magnification greatly dominates over Doppler magnification at small
inclination angles, but that the Doppler amplification is still prevalent for
a much wider range of inclination angles.

To determine for what systems a Doppler-boost detection will be plausible we
ask: when will the count rate be high enough for Poisson error bars on the
count number to be smaller than the maximum Doppler amplification. We use the
unlensed count rate from Eq. (\ref{Eq:Bkg_cnt}) and the Doppler amplification motived above
to determine the region of binary parameter space where
\begin{eqnarray}
\mbox{Poisson} \ \mbox{error} &\leq& \mbox{Doppler amplified counts} \nonumber \\
\sqrt{ \mathcal{C}_0 \frac{P_{\orb}}{2} } &\leq& \left( \frac{3-\alpha}{c}\frac{1}{1+q}\sqrt{\frac{GM}{a}} \right)\mathcal{C}_0 \frac{P_{\orb}}{2} .
\label{Eq:DopId}
\end{eqnarray}
The analogous criterion for the lensing case is nearly identical to the count rate criteria depicted by the teal region of Figure \ref{Fig:2}.

\begin{figure}
\begin{center}
\includegraphics[scale=0.33]{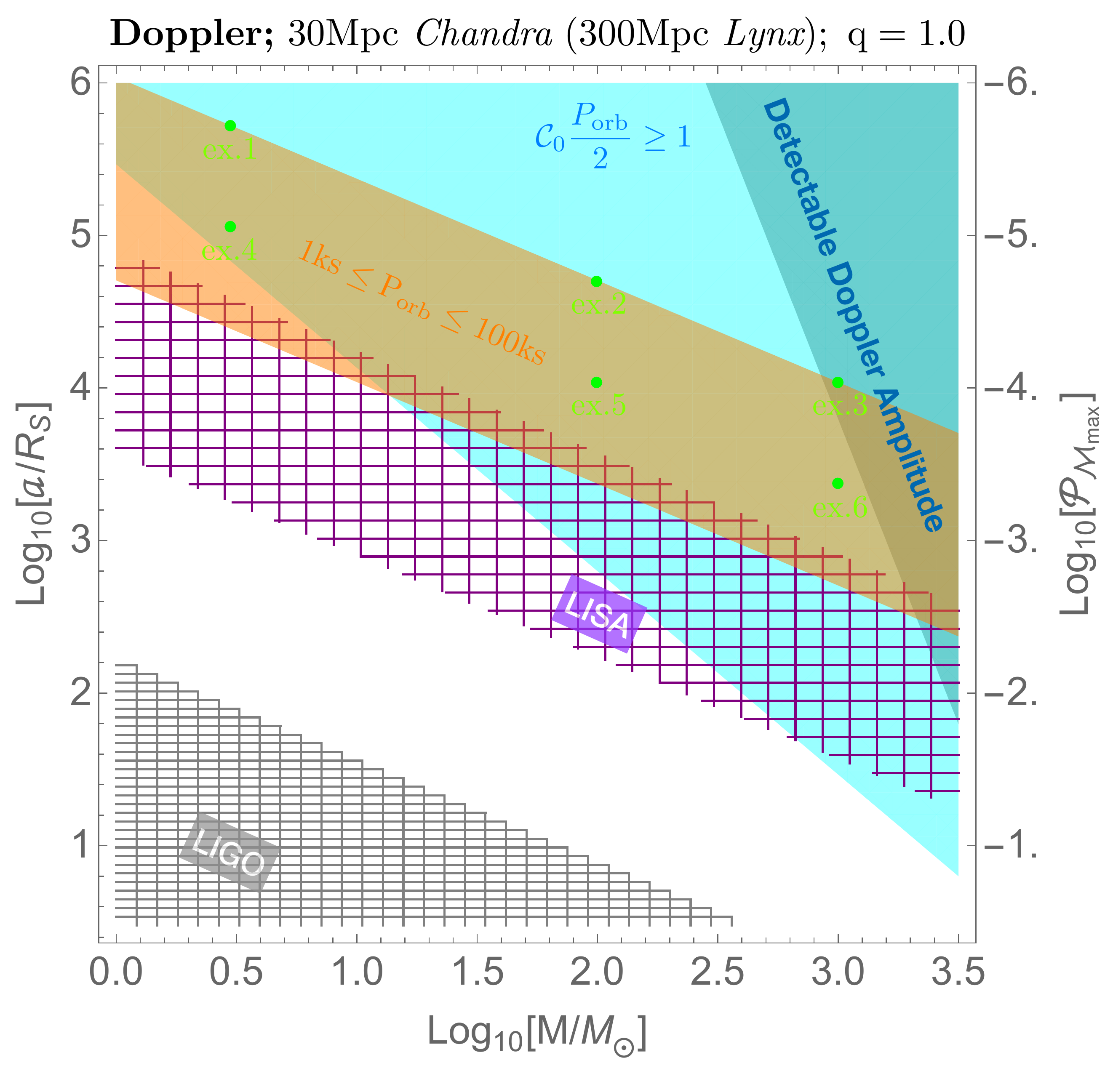}
\end{center}
\vspace{0pt}
\caption{
The same as Figure \ref{Fig:2}, except considering periodic modulation due
only to the Doppler boost. Here the \textcolor{cyan}{teal region} labeled
$\left<\mathcal{C}\right> P_{\orb}/2 \geq1$, denotes where the average flux
over half of an orbital period results in at least one count for a 
\Chandra-like (\Lynx-like) instrument and an Eddington luminosity source at 30~Mpc
(300~Mpc). The dark shaded region that overlays a subset of the teal region at
high binary masses denotes where there are enough counts to detect the Doppler
modulation, \textit{i.e.}, where the Poisson error in the number of counts is
smaller than the typical Doppler magnification, Eq. (\ref{Eq:DopId}).
}
\label{Fig:5Dop}
\end{figure}

In Figure \ref{Fig:5Dop} we plot this region in gray in a plot modeled after
Figure \ref{Fig:2}, but for the Doppler boost instead of lensing magnification.
As in Figure \ref{Fig:2}, the teal region specifies when the system is
detectable, except in the Doppler case we use the criteria that one count be
observed at the unlensed count rate over half of a binary orbit (rather than
requiring one count at the lensed count rate over the duration of the lensing event).
Importantly, the criteria for Doppler identification, Eq. (\ref{Eq:DopId}), is only
realized when many counts can be collected, for the most massive, longest
period systems. Figure \ref{Fig:5Dop} shows that Doppler-boost identification is
only viable for binaries with total mass greater than $\sim10^3\Msun$. In
these cases however, the detection probability can be much higher than for the lensing
case (a factor of $10-45\times$ higher for $M=10^3 \Msun$ and $P_{\orb} =
1-100$~ks). Whether or not there are enough BBH systems with mass above $10^3
\Msun$ within a few $\times 100$~Mpc to capitalize on the higher probability of finding
such Doppler-boost systems is uncertain.

\begin{figure}
\begin{center}
\includegraphics[scale=0.23]{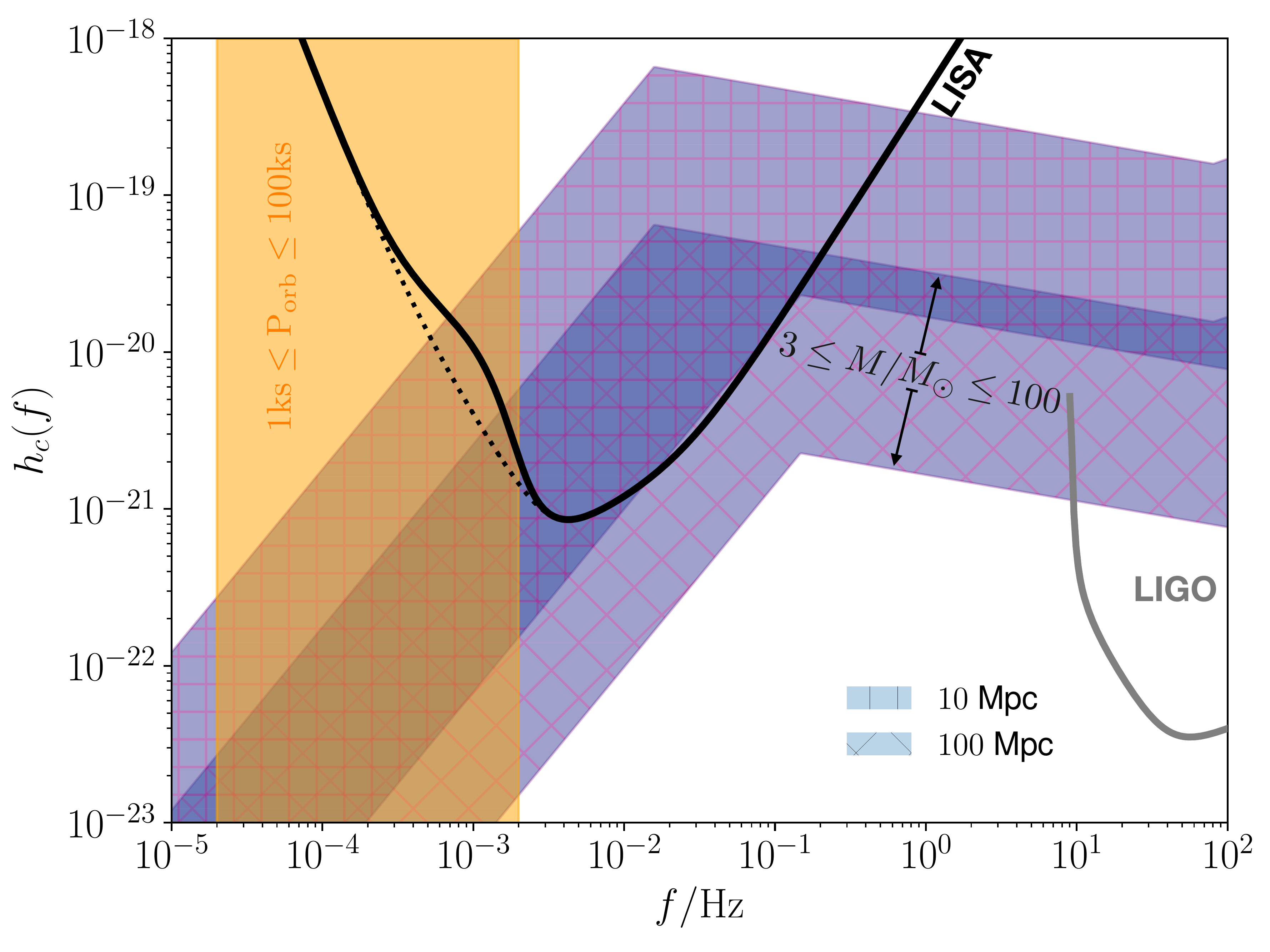} 
\end{center}
\vspace{0pt}
\caption{
The overlap of the orange region of Figures \ref{Fig:2} and \ref{Fig:5Dop} with the LISA band \citep{RobsonCornish:2018}. For reference we have shaded in purple the tracks of characteristic strain vs. GW frequency of CO binaries with masses ranging from $3-100 \Msun$ at distances of 10 and 100 Mpc as they pass through the LISA band (SNR=1 drawn in black) and the LIGO band (SNR=1 drawn in gray).
}
\label{Fig:6}
\end{figure}

While self-lensing offers a unique EM signature of CO binaries that may, in
their distant future, pass through the LISA band and merge in the LIGO band, it
may also offer an EM counterpart to low frequency, near-by LISA sources. This
is hinted at by the small overlap of the orange, teal, and hatched purple
regions in Figures \ref{Fig:2} and \ref{Fig:5Dop} and is explored in further
detail in Figure \ref{Fig:6}. Figure \ref{Fig:6} plots GW characteristic
strain vs. GW frequency, plotting the four-year-LISA (black; SNR=1) and O2
LIGO (grey; SNR=1) sensitivity curves for reference. The shaded purple regions
denote the GW strain over the lifetime of inspiralling CO binaries at 10 Mpc
(`+'' hatched) and 100 Mpc (`x' hatched) in the mass ranges of $3 \leq M/\Msun
\leq 100$ \citep[see][for an explanation of how these are drawn]{DSIII:2018}.
The orange region of Figures \ref{Fig:2} and \ref{Fig:5Dop} is also plotted in
orange in Figure \ref{Fig:6}, but in GW frequency space assuming circular
orbits. The overlap of orange and purple regions above the LISA sensitivity
curve shows that self-lensing EM counterparts to LISA sources are a
possibility for the closest and most massive binaries. This could be
interesting as a few such sources are predicted to exist
\citep{KremerMWLISA+2018}.

The shorter orbital period lensing events, below our $1$~ks cut (to the right
of the orange region in Figure \ref{Fig:6}), could still, in principle, be
detected. LISA inspirals should be monitored for such periodic EM emission.
Lensing events from sources entering the LIGO band are spaced too closely in
time and are too short in duration to be discerned with current or near future
X-ray telescopes. We note, however, that the microsecond cadence of gamma-ray
burst detectors such as Fermi GBM \citep{FERMIGBM:2015} may allow detection of
such events. The detection of an EM chirp from such a lensing
signature near merger has been investigated with a dynamical spacetime 
ray-tracing code by \citet{Schnittman:2018}.

\section{Conclusions}
\label{S:Conclusions}

We conclude with some caveats and prospects for future work. 

We have assumed circular binary orbits in our lensing models. Inclusion of
eccentricity changes the shape of lensing curves but does not significantly
alter the lensing duration or magnification.  Eccentric self-lensing is
explored in \citet{Spikey:2019}.

Here we have focused on X-ray observations of self-lensing because X-rays
naturally emanate from the smallest regions bound to BHs and NSs. However,
emission regions at optical and UV wavelengths can also exhibit self-lensing
signatures and may benefit from the upcoming and present all-sky time-domain
surveys such as the Zwicky Transient Facility \citep[ZTF,][]{ZTF} and The Large
Synoptic Survey Telescope \citep[LSST,][]{LSST}. That is, optical emission may
be more extended and, hence, not as highly magnified as the X-ray emission
region, but a much larger volume of space could be surveyed in optical. This
should be considered in future work.

Because of their repetition, symmetric pulse shape, and large magnifications,
self-lensing flares have high potential for being a uniquely identifiable
electromagnetic signature of compact-object binaries. Hence, in optical or in
X-ray, these unique signatures should be searched for. Null search results
will constrain the merging fraction in combination with the EM-bright lifetime
of compact-object binaries, while a detection of self-lensing offers a direct
window into their formation and early life.

\section*{Acknowledgements}
The authors thank the anonymous referee and also Jeff J.
Andrews, Deepto Chakrabarty, Pat Slane, Bradford Snios and Johan Samsing for
useful comments and discussions. Financial support was provided from NASA
through Einstein Postdoctoral Fellowship award number PF6-170151 (DJD) and
also through the Smithsonian Institution.

\bibliographystyle{mnras}
\bibliography{refs}
\end{document}